\documentclass[prd,preprint,showpacs,groupedaddress,fixfloat]{revtex4-1}
\usepackage{amsmath}
\usepackage{amsfonts}
\usepackage{amssymb}
\usepackage{geometry}
\usepackage{natbib}
\usepackage[english]{babel}
\usepackage{graphicx}
\usepackage{epstopdf}
\usepackage{subfigure}
\usepackage{caption}
\usepackage{multirow}
\usepackage{indentfirst}
\usepackage{mathrsfs}
\usepackage{chngcntr}
\usepackage{xcolor}
\usepackage{tikz}
\usetikzlibrary{patterns}
\usepackage[colorlinks,linkcolor=blue,anchorcolor=blue,citecolor=blue]{hyperref}

\definecolor{blau1}{rgb}{0,0.12,0.50}   % {  0, 31,127} blau1 ZARM
\definecolor{blau2}{rgb}{0.6,0.69,1}    % {153,177,255} blau2
\definecolor{blau3}{rgb}{0.87,0.90,1}   % {221,229,255} blau3
\definecolor{blau4}{rgb}{0,0.062,0.27}  % {  0, 16, 68} blau4
\definecolor{orange}{rgb}{1,0.88,0.50}  % {255,224,128} orange - blau1 invertiert
\definecolor{rot}{rgb}{0.79,0.31,0}     % {202, 78,  0} rot

\newcommand{\circpdash}[1]{%
\begin{tikzpicture}\draw[color=#1,dashed,dash pattern=on 1.5pt off 1.5pt] (0,0) circle (.27);\end{tikzpicture}}

\newcommand{\rectc}[1]{%
\begin{tikzpicture}\draw[fill=#1, color=#1]	(0,0) rectangle (.85,.4);\end{tikzpicture}}

\newcommand{\rectr}[1]{%
\begin{tikzpicture}%
\draw[color=#1, pattern=north east lines, pattern color=#1]	(0,0) rectangle (.85,.4);%
\end{tikzpicture}}
\newcommand{\rectl}[1]{%
\begin{tikzpicture}%
\draw[color=#1, pattern=vertical lines, pattern color=#1] (0,0) rectangle (.85,.4);%
\end{tikzpicture}}

\begin{document}

  \setlength{\parindent}{2em}
  \title{Shadow of topologically charged rotating braneworld black hole}
  \author{Hao-Ran Zhang} \author{Peng-Zhang He} \author{Lei-Shao} \author{Yuan Chen} 
  \author{Xian-Ru Hu} \email[Xian-Ru Hu: ]{huxianru@lzu.edu.cn}
  \affiliation{Lanzhou Center for Theoretical Physics $\&$ Research Center of Gravitation, Lanzhou University, Lanzhou 730000, China}
  \date{\today}

  \begin{abstract}
  
In this paper, we discuss optical properties of the topologically charged rotating black hole. We study the horizon, the photon region, the shadow of the black hole and other observables. The results show that in addition to the black hole spin parameter $a$, the other two parameters, tidal charge $\beta$ and electric charge $q$, are also found to affect the horizon, the photon region and the black hole shadow. In a certain range, with the increase of the three parameters, the horizon distance, shape of the photon region and the black hole shadow will all shrink. Moreover, with the increase of these three parameters, the distortion parameter $\delta_{s}$ gradually increases, while the peak of the black hole energy emission rate decreases.
  \end{abstract}

  \maketitle

\section{Introduction}
As a strange celestial body, the study of black holes has always attracted the attention of physicists. Recent years, through a project called Event Horizon Telescope (EHT), mankind got the first image of the shadow of the M87 supermassive black hole (SMBH) \cite{event2019first,akiyama2019first1,akiyama2019first2,akiyama2019first3,akiyama2019first4,akiyama2019first5}, which led us into a new era of black hole physics. It is foreseeable that with the improvement of the level of observation and the acquisition of more experimental data, we will have a deeper understanding of the nature of black holes, which will also provide scientists with feasible methods to test different gravitational theories.

In order to study the nature of black holes, we must determine the photon regions firstly. In Schwarzschild case, the black hole remains static, photons with critical angular momentum will form a spherical photon sphere near the black hole at this time \cite{teo2003spherical,perlick2004gravitational}, which is an area filled with null geodesics. But in Kerr case, the black hole is rotating, the spherical photon sphere is split into photon regions \cite{grenzebach2014photon,trimble2017shadow}, and these approximately circular null geodesics describe the boundary of the black hole shadow. So the nature of photon regions take great significance to the research of black hole shadow.

The black hole shadow is one of the important conclusions of the general gravitational theory. It is formed by the null geodesics in the strong gravitational region near the black hole. A photon with a large angular momentum flying from infinity, due to the influence of the strong gravitational force of the black hole, the path of motion will be deflected and eventually fly to infinity. There is evidence that a massive black hole exists in the center of the galaxy. Since the galaxies are rotating, it is very likely that the black hole is also rotating. Spin and mass are two important parameters for studying the properties of black holes, and black hole shadow is one of the effective methods to observe the spin and mass of black holes. For the non-rotating Schwarzschild black hole, its shadow was first studied by Synge and
Luminet \cite{synge1966escape,luminet1979image}. The black hole shadow in Schwarzschild case \cite{darwin1959gravity,ohanian1987black,nemiroff1993visual,bozza2001strong}, the rotating black hole with electric charge \cite{grenzebach2014photon} or gravitomagnetic and other spherically symmetric black holes have been extensively studied. It has been demonstrated that the black hole shadow in Schwarzschild case is a perfect circle \cite{bozza2010gravitational} in vacuum \cite{virbhadra2009relativistic} or in plasma \cite{morozova2013gravitational,tsupko2013gravitational}. But in rotating case, since the black hole has angular momentum, the shadow will be deformed \cite{falcke1999viewing,chandrasekhar1998mathematical,nedkova2013shadow,bambi2009apparent,zakharov2005measuring,zakharov2005direct,de2011estimating,de2000apparent}. Shadows of Kerr-Newman black holes were obtained in \cite{de2000apparent}, while the naked singularities with deformation parameters was discussed in \cite{hioki2009measurement}, multi-black holes \cite{yumoto2012shadows} and Kerr-Taub-NUT black hole in \cite{abdujabbarov2013shadow}. These works are also used to study other black hole or wormhole backgrounds \cite{wei2020testing,zhang2020optical,chen2020optical,hioki2008hidden,amarilla2010null,atamurotov2013shadow,papnoi2014shadow,shaikh2018shadows,abdolrahimi2015distorted,atamurotov2015optical}.

In order to be able to compare observations with theoretical predictions, we need to consider the corresponding observables. Since the shape of the shadow of a black hole depends on its boundary, we can study these observables through the characteristic points on the boundary. In \cite{hioki2009measurement}, the authors introduced a series of observables to describe the shadow shape of the Kerr black hole or naked singularity. Based on these observations, the oblique angle of the observer and the spin of the black hole can be well described. And in \cite{abdujabbarov2015coordinate}, the authors introduces a series of shadow distortion parameters based on a new form of describing black hole shadows.

There is no clear evidence that black holes cannot exist in higher dimensions, and in dimensions higher than four, there are more degrees of freedom, so the uniqueness theorems do not hold. And the braneworld is an interesting high-dimensional cosmological model \cite{langlois2002brane,brax2003cosmology,maartens2010brane}, which motivated by string theory (M-theory). The braneworld model has been proposed in which the standard fields are confined to a four-dimensional (4D) world viewed as a hypersurface (the brane) embedded in a higher-dimensional space-time (the bulk) through which only gravity can propagate \cite{rubakov2001large,csaki2004tasi}. The apparent shape of rotating braneworld black holes were investigated in \cite{amarilla2012shadow,schee2009optical}. And in \cite{larranaga2013topologically}, the authors studied the topologically charged rotating braneworld black hole. This black hole has two types of charge, one arising from the bulk Weyl tensor and one from the gauge field trapped on the brane, which will cause interesting results.

This paper is organized as follows. In section \ref{II}, we briefly introduced the solution of the topologically charged rotating black hole and discussed the horizon. The geodesic equations and orbital equations of photons are given in section \ref{3}. Further, we drew the photon region of the charged topological black hole and the black hole shadow in the section \ref{4} and section \ref{5} respectively. In addition, we also discussed the deformation of the black hole shadow and studied the energy emissivity. Finally, we summarize the work of the article in the section \ref{6}.

\section{topologically charged rotating braneworld black hole}\label{II}
  The metric of a topologically charged rotating black hole in the Boyer-Lindquist coordinates from \cite{larranaga2013topologically} reads 
   \begin{equation}\label{dg}
   \begin{split}
   \mathrm{d}s^{2}=\frac{\Delta-a^2\sin^2{\theta}}{\Sigma}&\mathrm{d}t^{2}-\frac{\Sigma}{\Delta}\mathrm{d}r^{2}-\Sigma\mathrm{d}\theta^{2}+2a\sin^2{\theta}\left(1-\frac{\Delta-a^2\sin^2{\theta}}{\Sigma}\right)\mathrm{d}t\mathrm{d}\varphi-\\ &\sin^2{\theta}\left[\Sigma+a^2\sin^2{\theta}\left(2-\frac{\Delta-a^2\sin^2{\theta}}{\Sigma}\right)\right]\mathrm{d}\varphi^2,
  \end{split}
  \end{equation}
 with $\Delta$ and $\Sigma$ given by
  \begin{equation}
  \Delta=r^{2}-2 M r+a^{2}+Q\left(r\right),
  \end{equation}
  \begin{equation}
  \Sigma=r^{2}+a^2\cos^2{\theta},
  \end{equation}
and
  \begin{equation}
  Q\left(r\right)=\beta+q^2+\frac{\kappa_{5}^{\,4}q^4}{240r^4},
  \end{equation}
here $a$ is spin parameter, $\beta$ is tidal charge, $q$ is the electric charge, and $\kappa_{5}^{\,4}$ is the five-dimensional gravity coupling constant. For the sake of calculation, we have chosen $\kappa_{5}^{\,4}=1$. The solution reduce to Schwarzschild solution when $a=\beta=q=0$ and $\kappa_{5}^{\,4}=0$.

As we can see, the metric is singular when 
  \begin{equation}\label{eq:radii}
  \Delta=r^{2}-2 M r+a^{2}+\beta+q^2+\frac{q^4}{240r^4}=0,
  \end{equation}
it can be seen from the above equation, the radii of horizons depend on the rotation parameter $a$ and the tidal charge $\beta$. Obviously, this equation is unable to solved. However, the numerical analysis of Eq.~(\ref{eq:radii}) suggests the possibility of two roots of a set of values of parameters, which correspond the inner horizon $r_{-}$(smaller root) and the outer horizon $r_{+}$ (lager root), respectively. The variation of $\Delta$ with respect to $r$ for the different values.
\par
\begin{figure}[htbp]
	\centering
	\includegraphics[width=.47\textwidth]{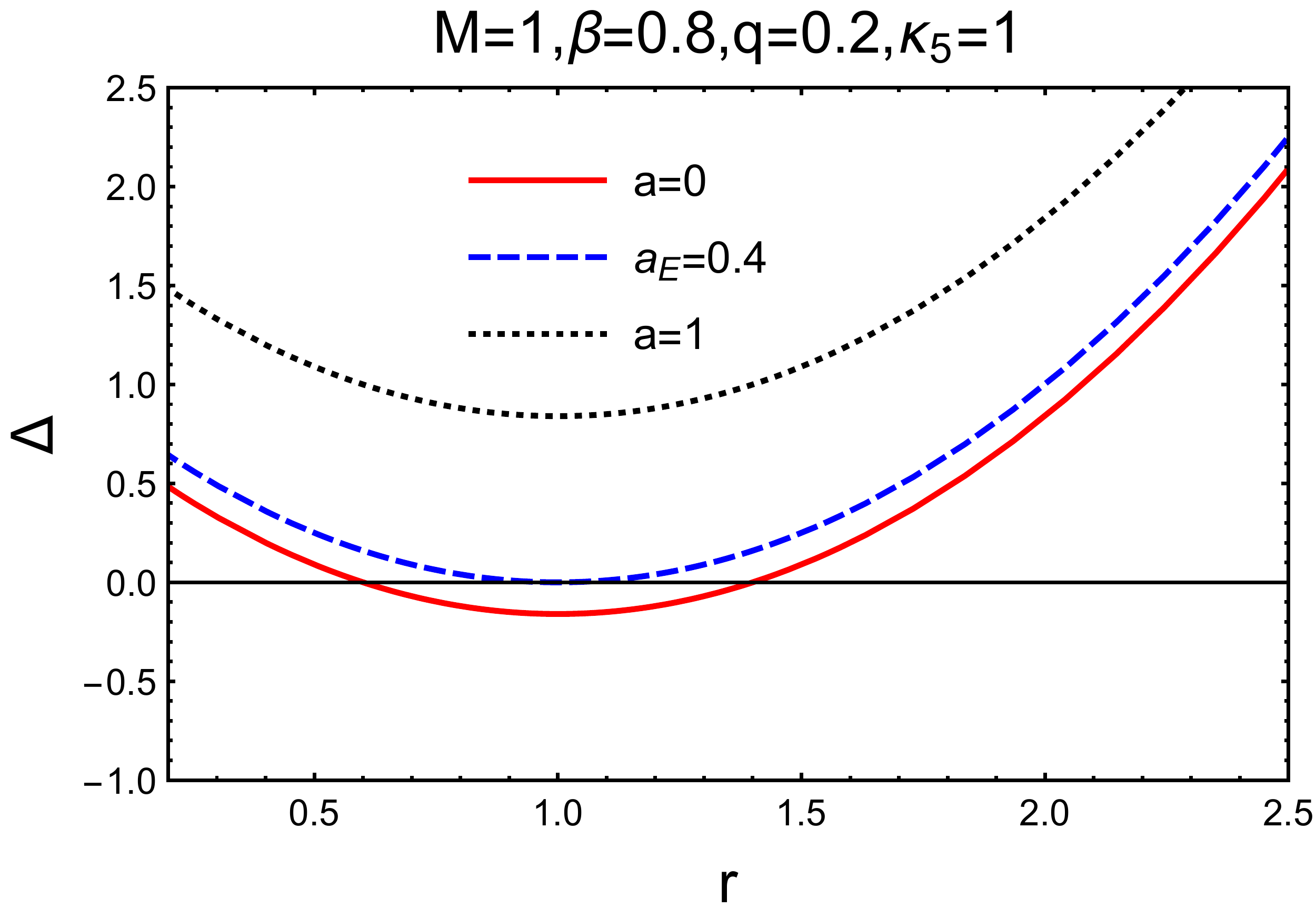}
	\includegraphics[width=.47\textwidth]{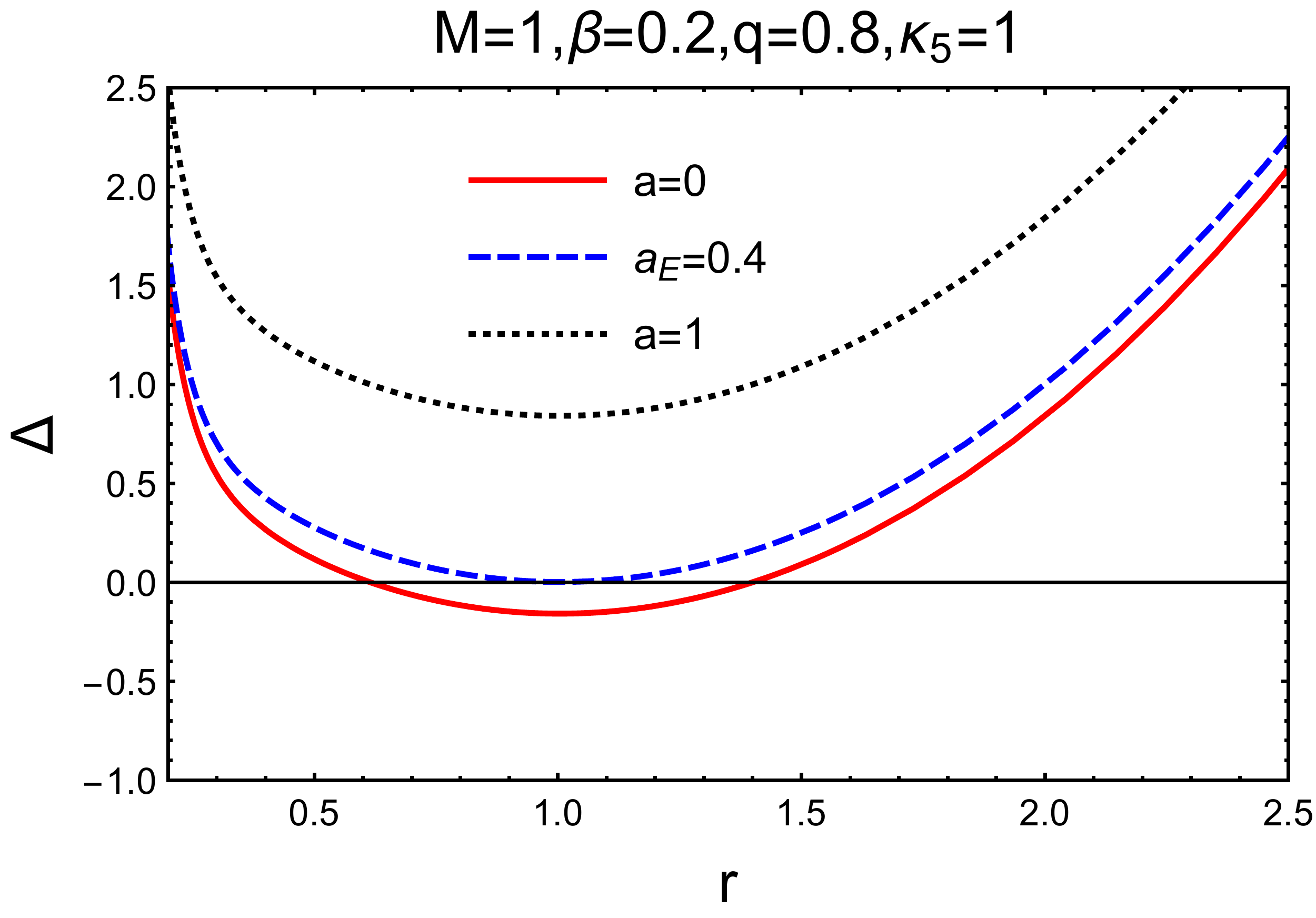}
	\caption{Plot showing the behavior of horizons vs. $r$ for $a$ set of fixed values of $M=\kappa_{5}=1$, $\beta$ and $q$ by varying $a$.}
	\label{horizon1}
\end{figure}

  The variation of $\Delta$ with respect to $r$ for different values of parameters $a$, with fixed $\beta$ and $q$ is depicted in Fig. \ref{horizon1}. As can be seen from Fig. \ref{horizon1}, for fixed parameters $\beta$ and $q$, the radii of outer horizon decrease with the increasing a while the radii of inner horizon increase when $a<a_{E}$. By calculating $\Delta=0$ and $\mathrm{d}\Delta/\mathrm{d}r=0$, we can obtain the critical rotation parameter $a_{E}$ and the corresponding critical radius $r_{E}$. Eq.~(\ref{eq:radii}) will have no root if $a>a_{E}$, inner horizon and outer horizon will disappear, in other words, the black hole will not exists any more.
\par
\begin{figure}[htbp]
	\centering
	\includegraphics[width=.47\textwidth]{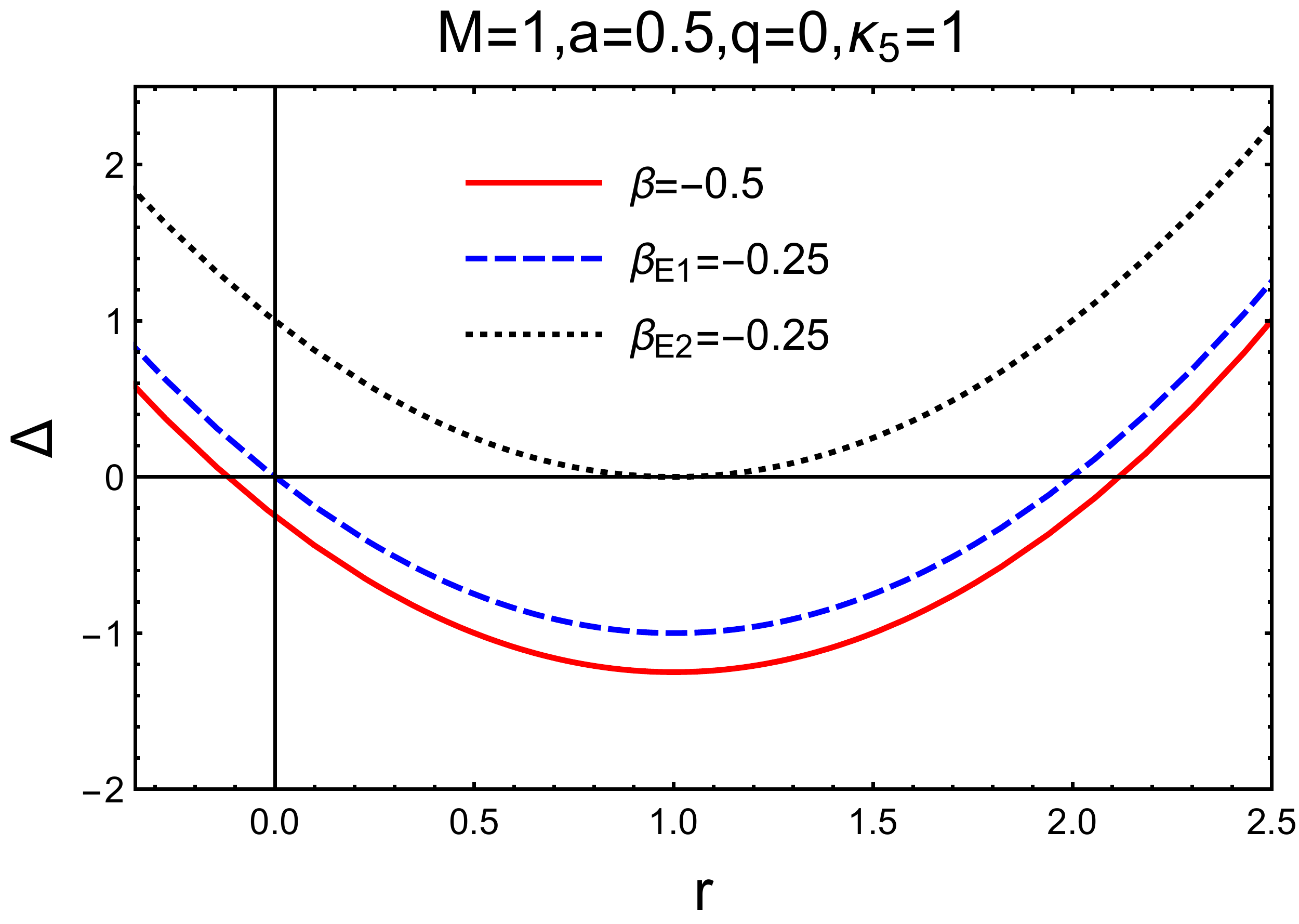}
	\includegraphics[width=.47\textwidth]{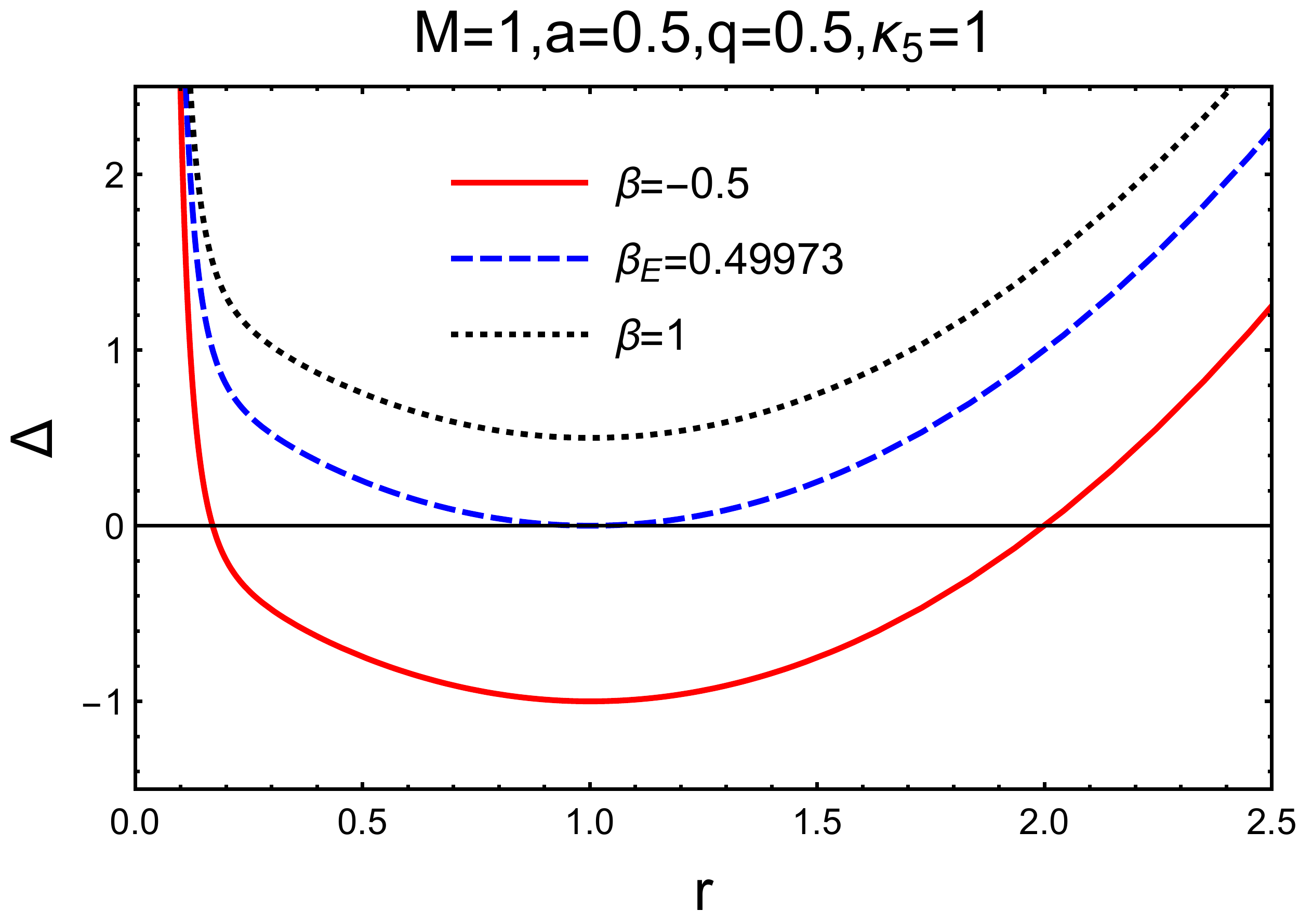}
	\caption{Plot showing the behavior of horizons vs. $r$ for $a$ set of fixed values of $M=\kappa_{5}=1$, $a$ and $q$ by varying $\beta$}
	\label{horizon2}
\end{figure}
\begin{figure}[htbp]
	\centering
	\includegraphics[width=.47\textwidth]{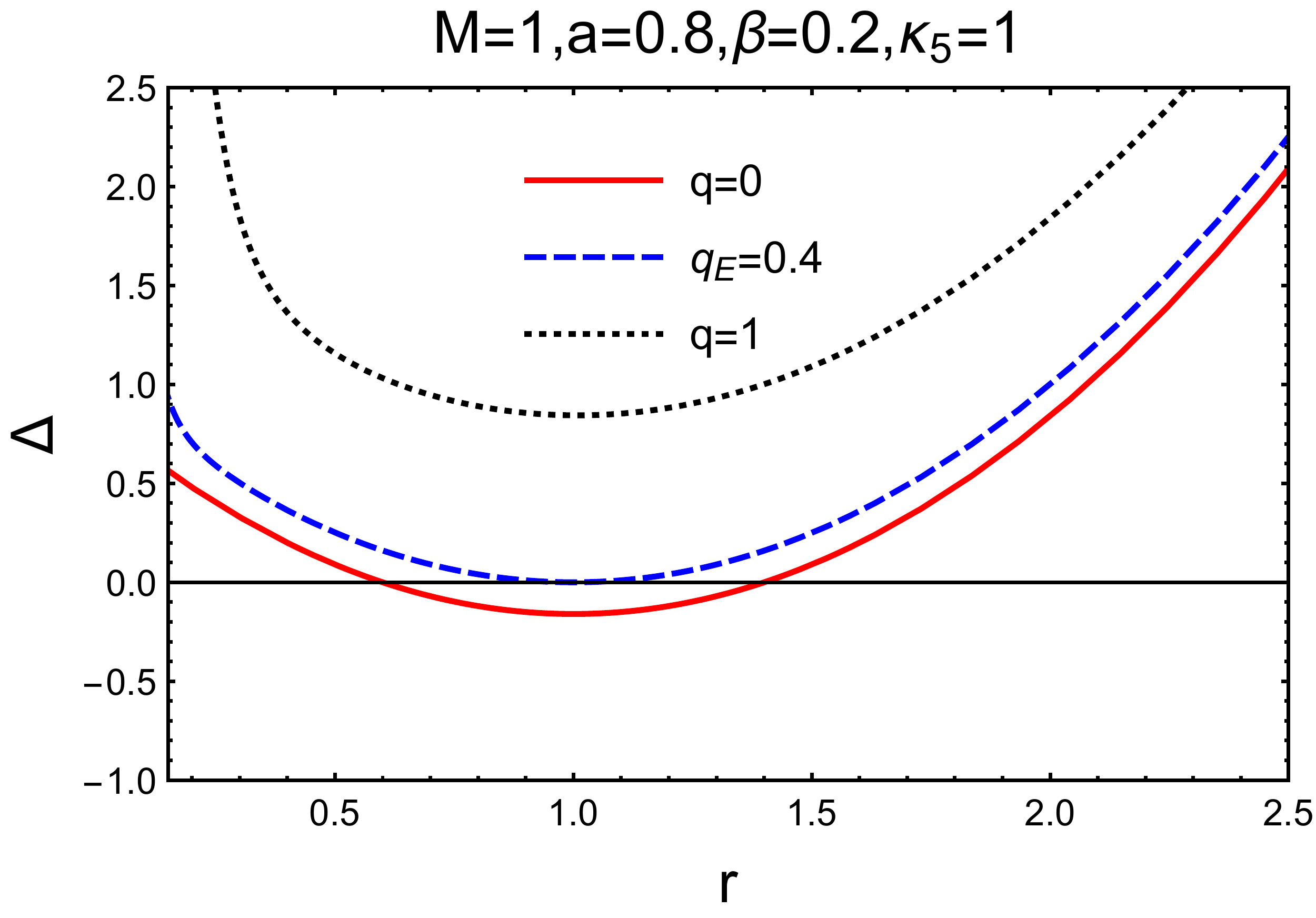}
	\includegraphics[width=.47\textwidth]{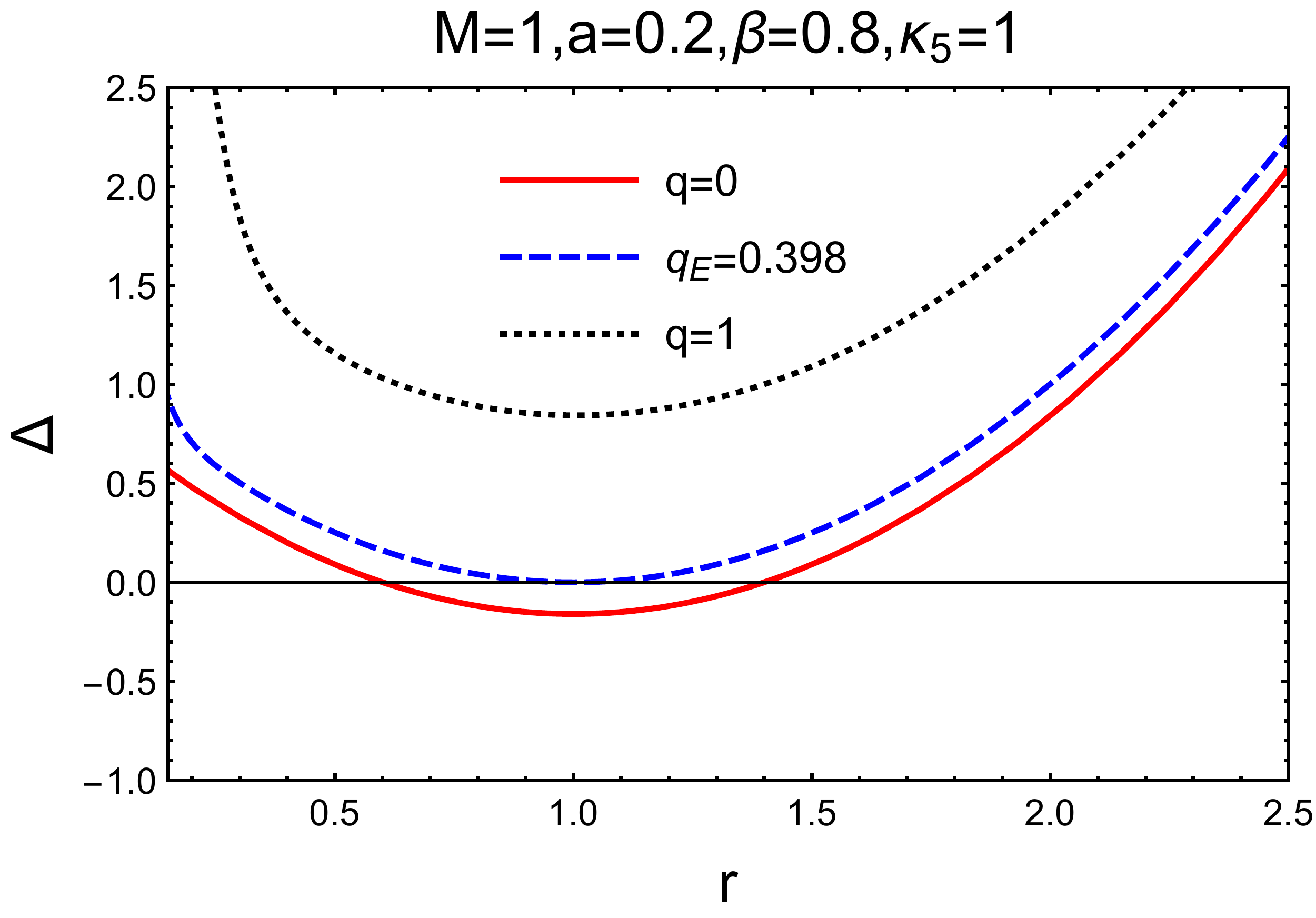}
	\caption{Plot showing the behavior of horizons vs. $r$ for $a$ set of fixed values of $M=\kappa_{5}=1$, $a$ and $\beta$ by varying $q$}
	\label{horizon3}
\end{figure}

  We can analyze the other two parameters, $\beta$ and $q$, in the same way. And we can see the variation of $\Delta$ with respect to $r$ for different values of parameters $\beta$ (or $q$) with fixed $a$ in Figs. \ref{horizon2} and \ref{horizon3}, respectively. Similar to changing parameter $a$, as $\beta$ (or $q$) increases, the distance between two horizons gradually decreases, when $\beta=\beta_{E}$ (or $q=q_{E}$), two horizons overlap and horizon will disappear if $\beta>\beta_{E}$ (or $q>q_{E}$). There is no evidence that $\beta$ cannot take a negative value. Interestingly, as the value of $\beta$ decreases, the inner horizon will gradually approach the origin, and eventually cross the origin and reach the $r<0$ area. However, if $q\neq0$, the inner horizon will always be greater than zero.


\section{photon orbits}\label{3}
  In this section, we will give a brief introduction to the photon orbits in the topologically charged rotating braneworld black hole. To study the equation of motion of photons in the field of a topologically charged rotating braneworld black hole, we begin with the Lagrangian which reads
    \begin{equation}\label{lglr}
  L=\frac{1}{2}g_{\mu\nu}\dot{x}^{\mu}\dot{x}^{\nu},
    \end{equation}
where an overdot denotes the partial derivative with respect to an affine parameter. In order to analyze the general orbit of photons around a black hole, we study the separability of the Hamilton-Jacobi equation for which we adopt the approach originally suggested by Carter \cite{carter1968global}. The Hamilton-Jacobi equation in braneworld black hole space-time~(\ref{dg}) with the metric
tensor $g_{\mu\nu}$ takes the following general form:
  \begin{equation}\label{HJQ}
  \frac{\partial{S}}{\partial{\lambda}}=-\frac{1}{2}g^{\mu\nu}\frac{\partial{S}}{\partial{x}^{\mu}}\frac{\partial{S}}{\partial{x}^{\nu}},
  \end{equation}  
 where $\lambda$ is affine parameter and the action $S$ which can be decomposed as a sum:
  \begin{equation}\label{HJS}
  S=\frac{1}{2}m^{2}\lambda-Et+L_{\phi}\phi+S_{r}\left(r\right)+S_{\theta}\left(\theta\right),
  \end{equation}    
  where $E$ and $L_{\phi}$ are, respectively, the energy and the angular momentum in the direction of the axis of symmetry. Substituting~(\ref{HJS}) into~(\ref{HJQ}), one can obtain the equations of motion
  \begin{equation}
  \Sigma\frac{\mathrm{d}t}{\mathrm{d}\lambda}=\frac{-E{\Sigma}^{2}+a L_{\phi}\left(-\Delta+\Sigma\right)+a^2 \left(aL_{\phi}+E\left(\Delta-2\Sigma\right)\right)\sin^2{\theta}-a^{4}E\sin^{4}{\theta}}{\Delta},
  \end{equation}
  \begin{equation}
  \Sigma\frac{\mathrm{d}\phi}{\mathrm{d}\lambda}=-\frac{L_{\phi}}{\sin^2{\theta}}+\frac{a\left(aL_{\phi}+E\Delta-E\Sigma-a^{2}E\sin^2{\theta}\right)}{\Delta},
  \end{equation}
  \begin{equation}\label{r}
  \Sigma\frac{\mathrm{d}r}{\mathrm{d}\lambda}=\sqrt{\mathcal{R}},
  \end{equation}
  \begin{equation}
  \Sigma\frac{\mathrm{d}\theta}{\mathrm{d}\lambda}=\sqrt{\Theta},
  \end{equation}
  where $\mathcal{R}$ and $\Theta$ are given by
  \begin{equation}\label{R}
  \mathcal{R}=-\mathcal{K}\Delta-\Delta\left(L_{\phi}-aE\right)^2+\left(aL_{\phi}-\left(a^2+r^2\right)E\right)^2,
  \end{equation}
  \begin{equation}\label{Theta}
  \Theta={\mathcal{K}}+L_{\phi}^2+a^{2}E^{2}\cos^2{\theta}-\frac{L_{\phi}^2}{\sin^2{\theta}} ,
  \end{equation}
 with $\mathcal{K}$ denoting the Carter constant. Since the spacetime is asymptotically flat, the photon path is a straight line at infinity. However, when there a black hole is placed between the observer and the light source, the light will reach the observer after deflecting due to the strong gravitational field of the black hole. And now, we are going to discuss the radial motion of photon.
\par
The motion of photon is determined by two impact parameters $\xi=L_{\phi}/E$ and $\eta=\mathcal{K}/E$. From Eq.~(\ref{r}), we can obtain the circular photon orbits, which are very useful on determining the shape of black hole shadow. The conditions for these orbits are
  \begin{equation}\label{R0}
  \mathcal{R}=0, \frac{\mathrm{d}\mathcal{R}}{\mathrm{d}r}=0,
  \end{equation}
  from above equations, we get
  \begin{equation}\label{xi}
\xi=a+\frac{r^2}{a}-\frac{4r\Delta}{a\Delta'},
  \end{equation}
  \begin{equation}\label{eta}
\eta=\frac{r^2\left(-16\Delta^2-r^2\Delta'^2+8\Delta\left(2a^2+r\Delta'\right)\right)}{a^2\Delta'^2},
  \end{equation}
 these two parameters will help us to find the photon regions and the boundary of the black hole shadow, which will discuss in following sections.

\section{photon regions}\label{4}
\par
Now, we are interested in spherical null geodesics, i.e., null geodesics that stay on a sphere $r=$constant. Inserting these expression~(\ref{xi}) and~(\ref{eta}) into~(\ref{Theta}), which gives us an inequality that calculation the photon region $\mathscr{K}$:
\begin{equation}
\mathscr{K}:16a^2 r^2 \Delta \sin^2\theta\ge\left(4r\Delta-\Sigma\Delta'\right).
\end{equation}

A spherical null geodesic at $r=r_{p}$ is unstable with respect to radial perturbations if $R''({r_{p})}>0$, and stable if $R''({r_{p})}<0$. From Eq.~(\ref{R}), we can calculate the second derivative $R''$
\begin{equation}
\frac{\mathcal{R''}}{8E^2}\Delta'^2=2r\Delta\Delta'+r^2\Delta'^2-2r^2\Delta\Delta''.
\end{equation}

\begin{figure}[htbp]
	\centering
	\subfigure[$\beta$=q=0]{
		\begin{minipage}[t]{0.311\textwidth}
			\includegraphics[width=1\textwidth]{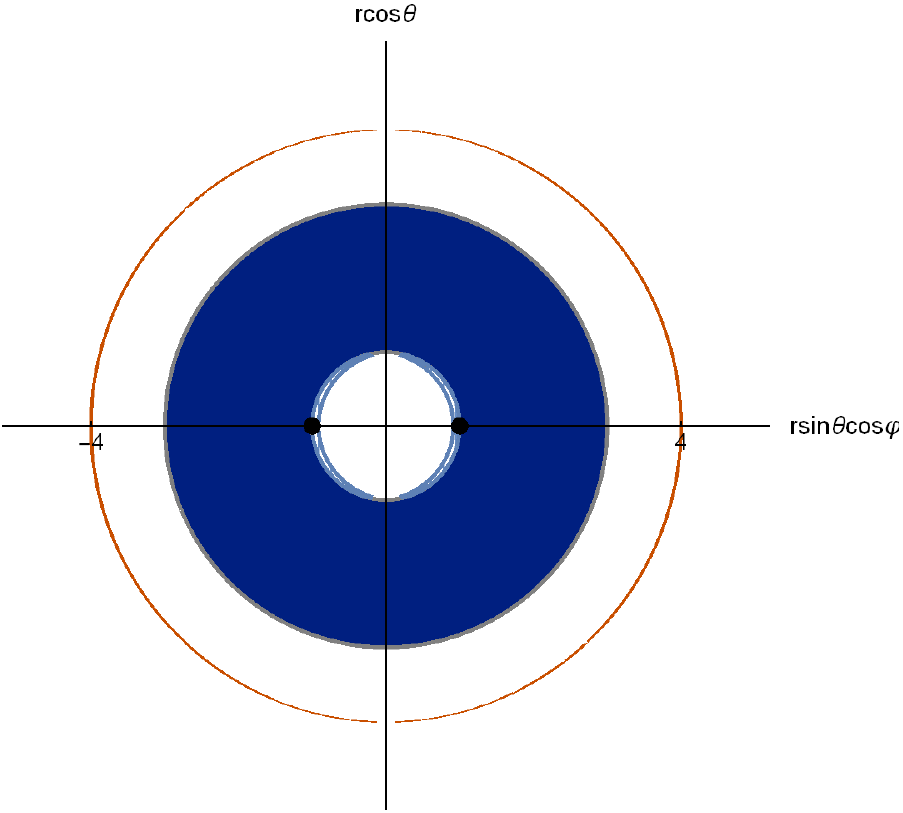}
			\includegraphics[width=1\textwidth]{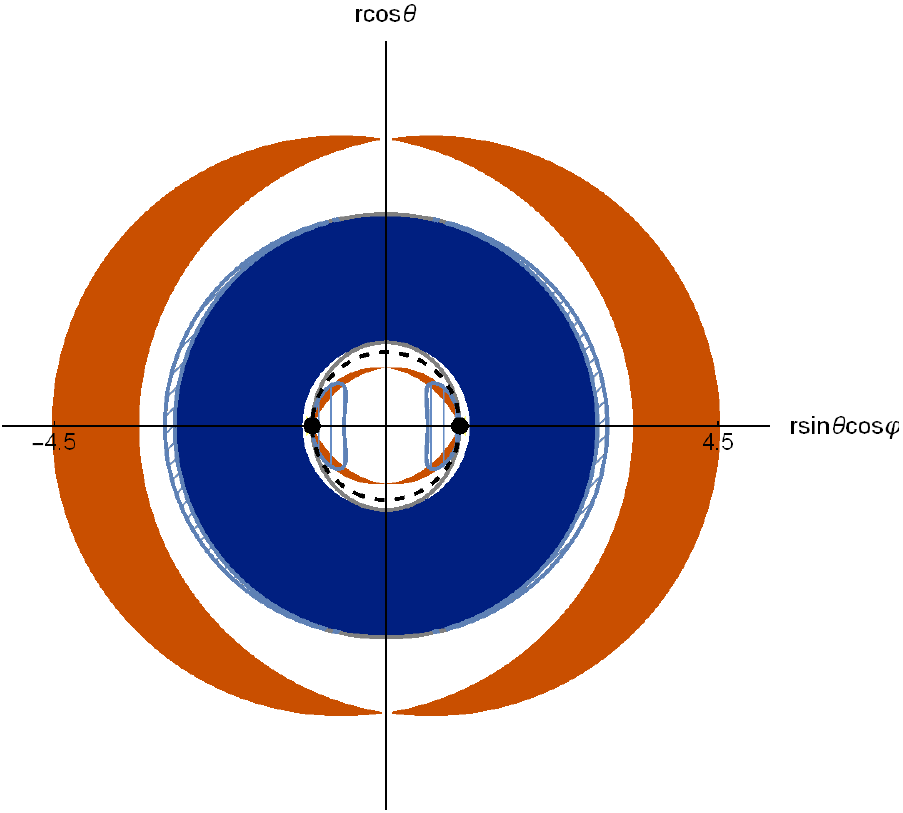}
			\includegraphics[width=1\textwidth]{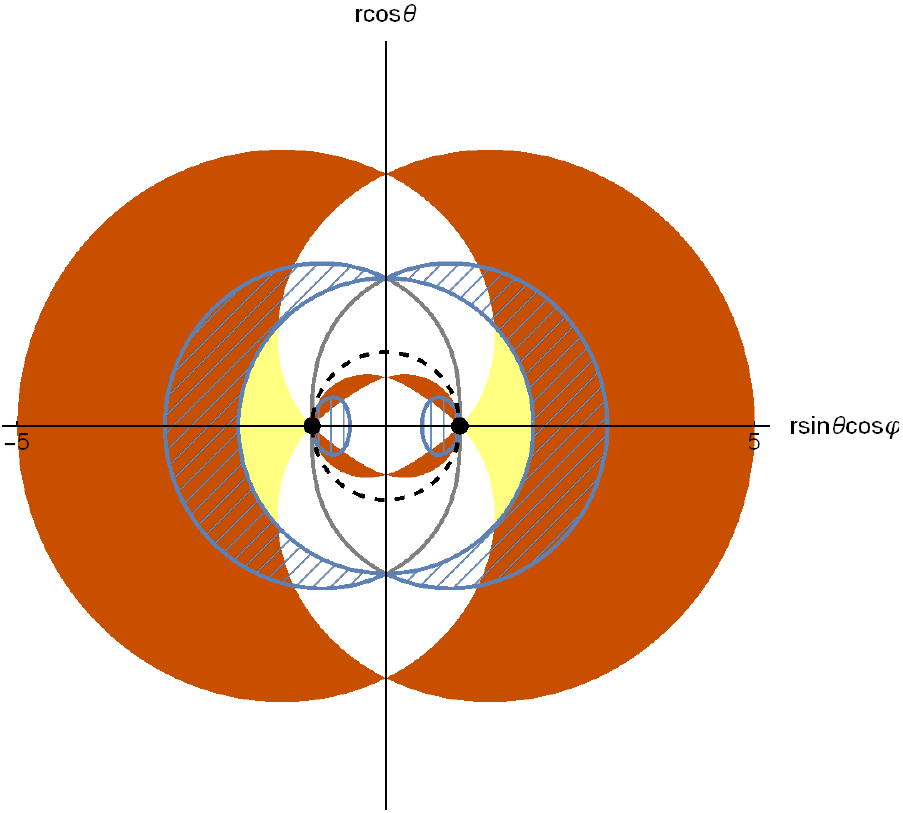}
	\end{minipage}}
	\subfigure[$\beta$=q=0.1]{
		\begin{minipage}[t]{0.311\textwidth}
			\includegraphics[width=1\textwidth]{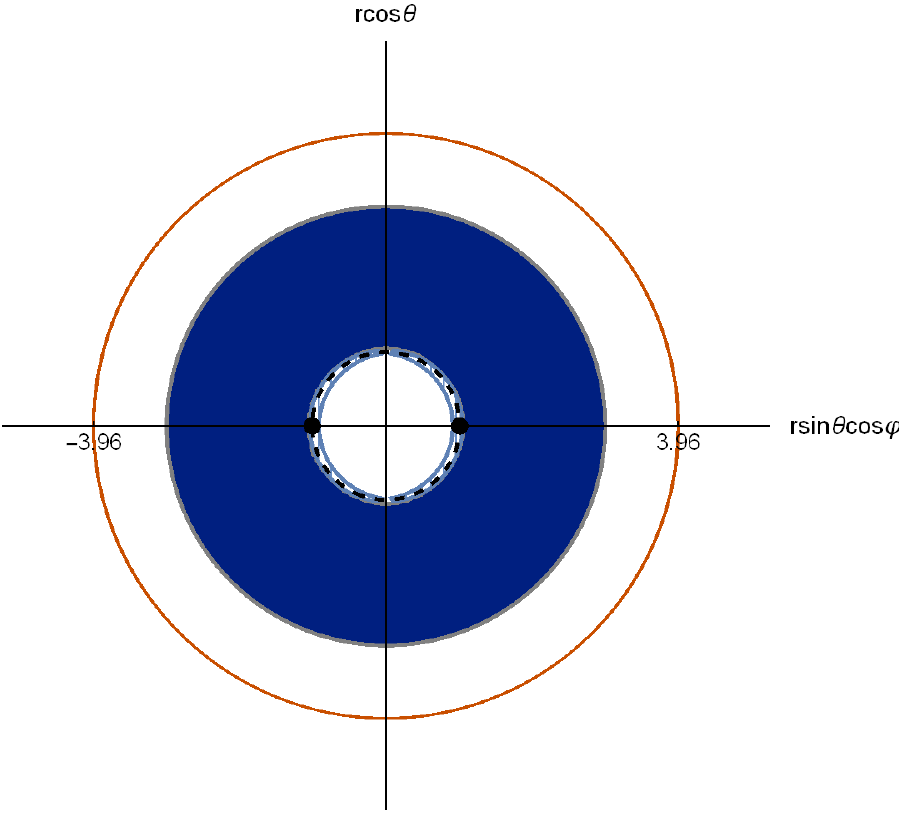}
			\includegraphics[width=1\textwidth]{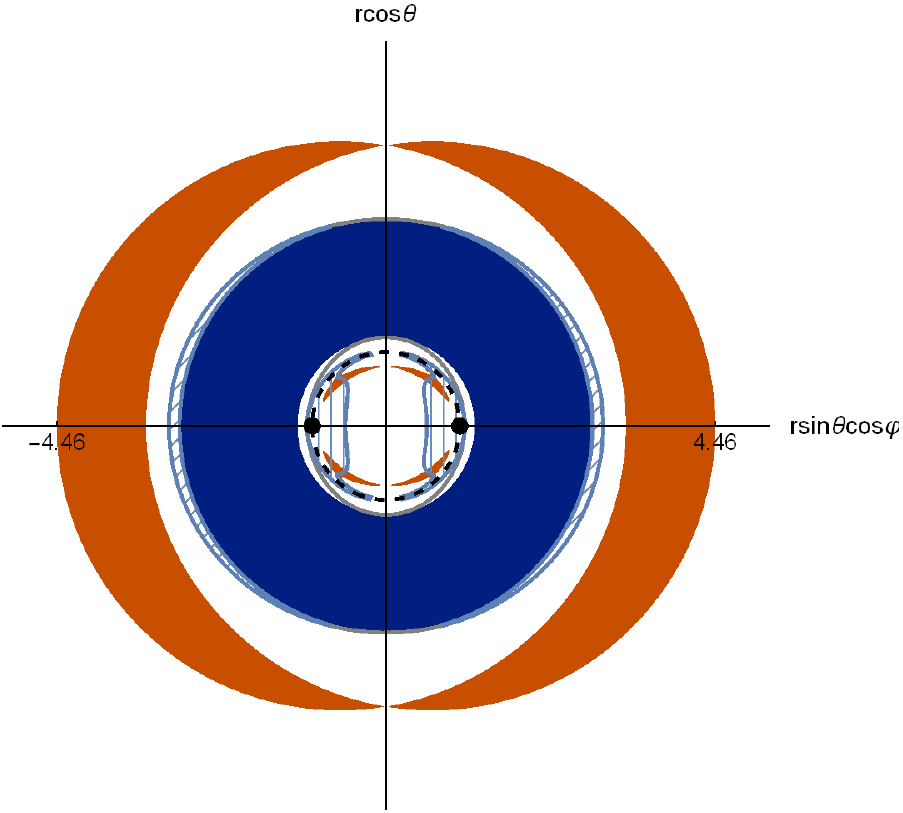}
			\includegraphics[width=1\textwidth]{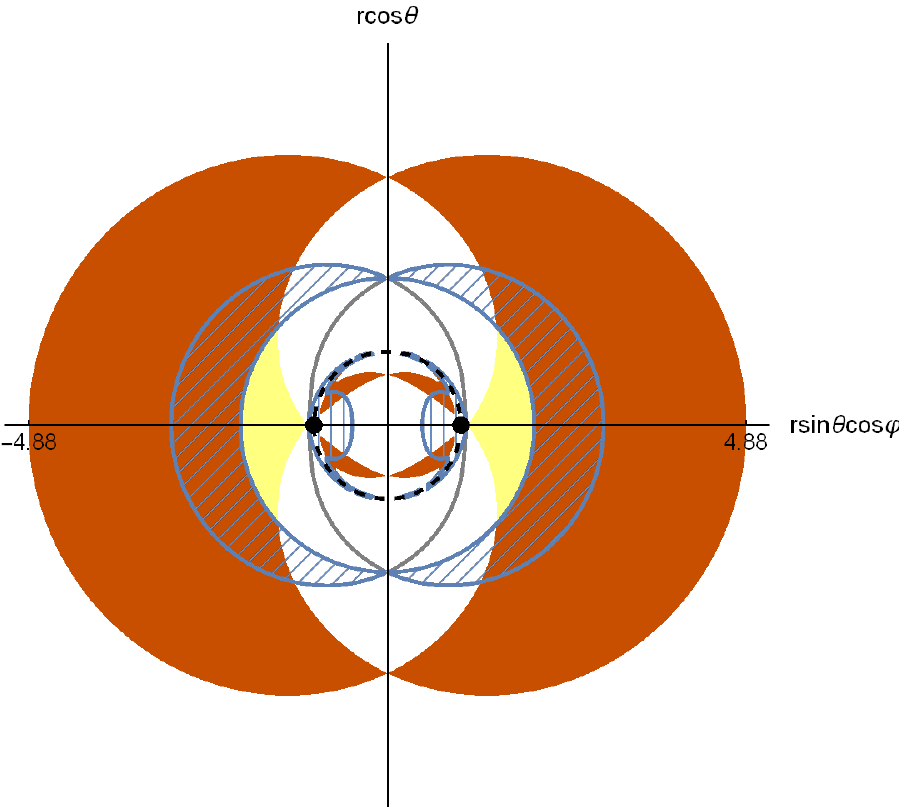}
	\end{minipage}}
	\subfigure[$\beta$=q=0.4]{
		\begin{minipage}[t]{0.311\textwidth}
			\includegraphics[width=1\textwidth]{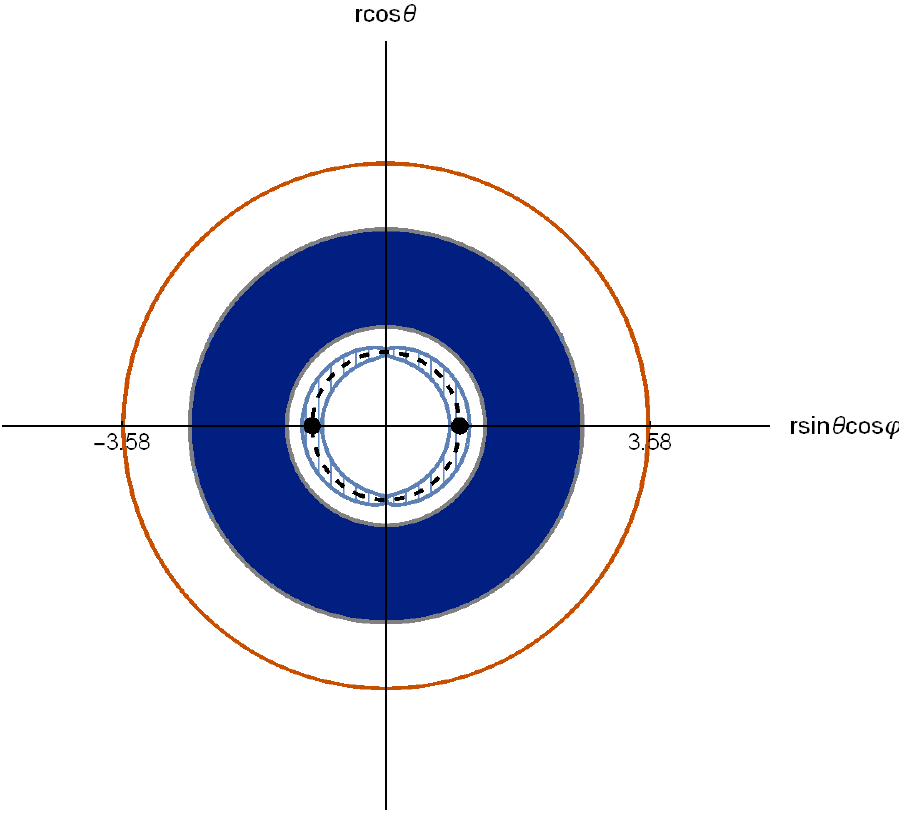}
			\includegraphics[width=1\textwidth]{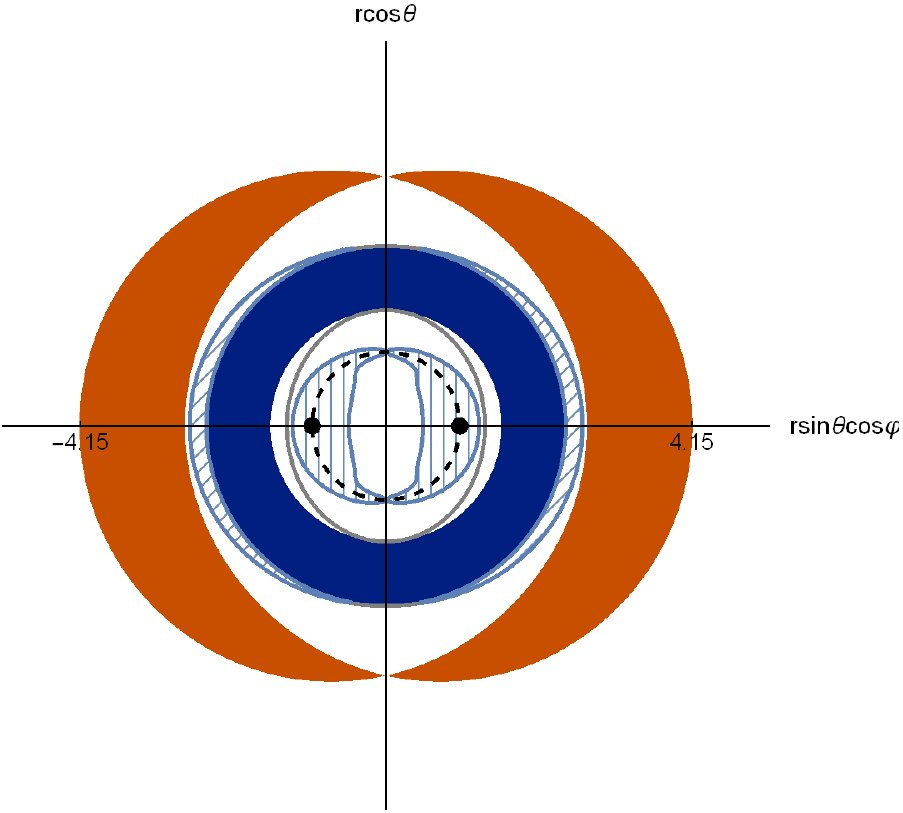}
			\includegraphics[width=1\textwidth]{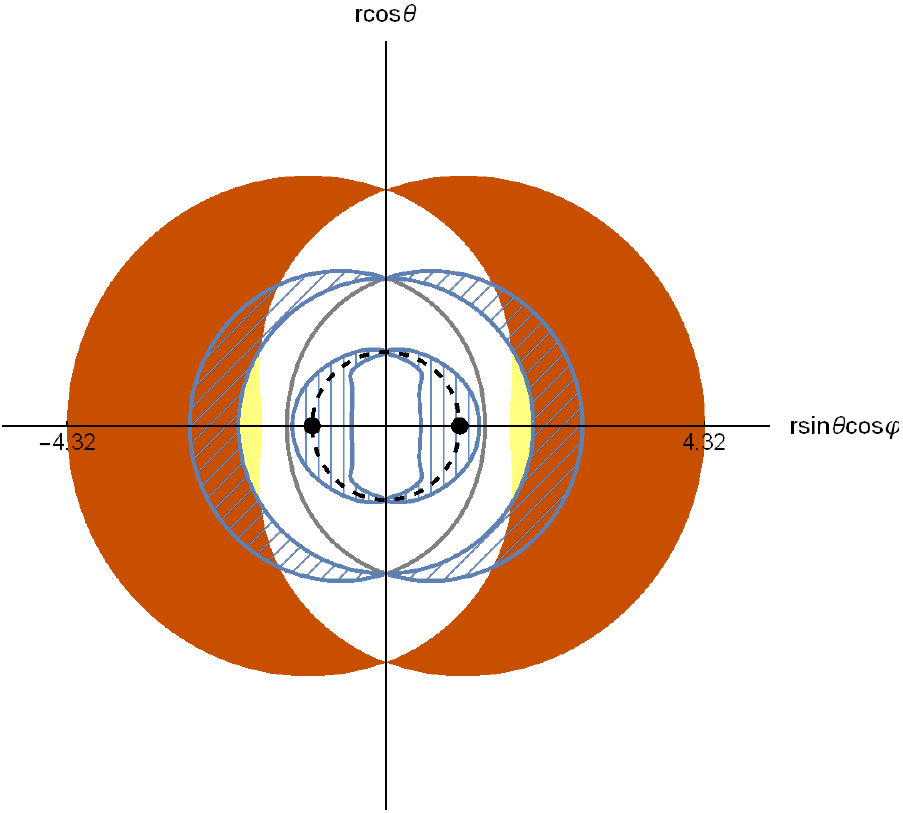}
	\end{minipage}}
	\caption{The shapes of the photon regions for different values of the parameters. The first line, $a=0.02$. The second line, $a=0.5$. The third line, $a=a_{Max}$.}
	\label{photon-1}
\end{figure}

\begin{figure}[htbp]
\begin{minipage}[t]{0.45\textwidth}
\includegraphics[width=1\textwidth]{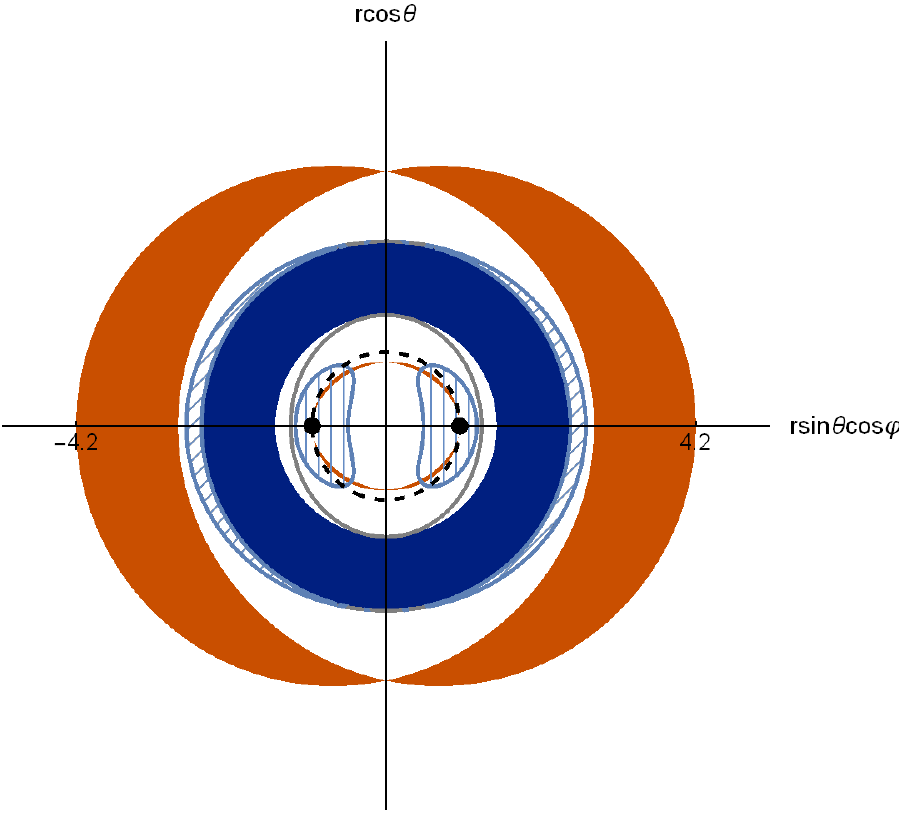}
\includegraphics[width=1\textwidth]{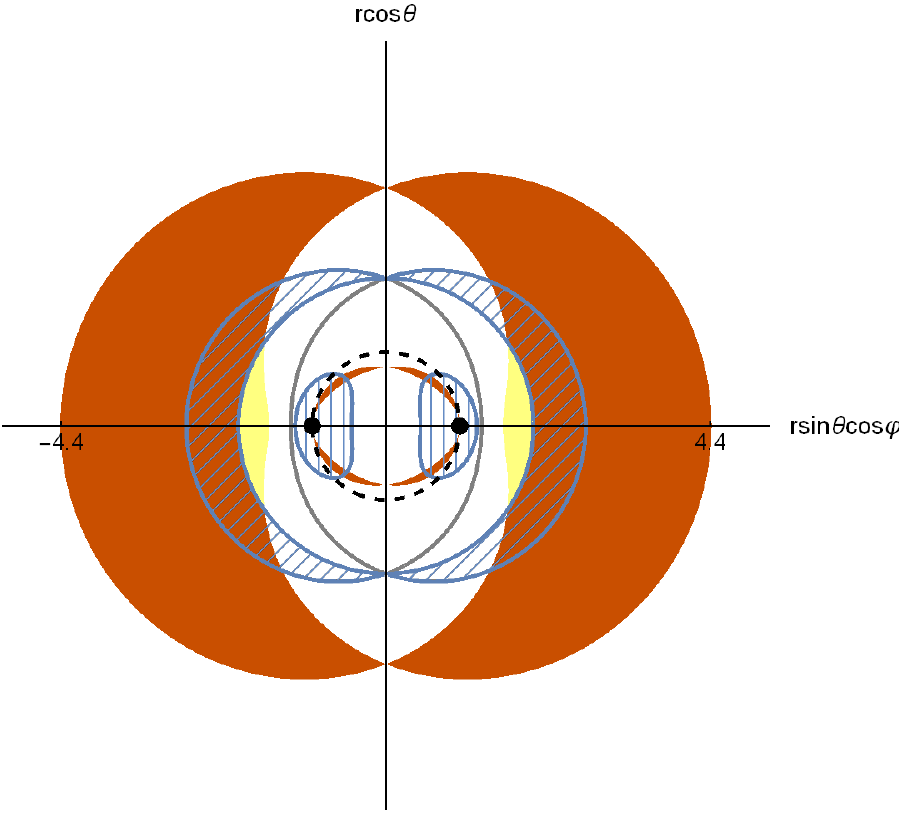}
\centerline{(a)$\beta$=0.2,q=0}
\end{minipage}
\begin{minipage}[t]{0.45\textwidth}
\includegraphics[width=1\textwidth]{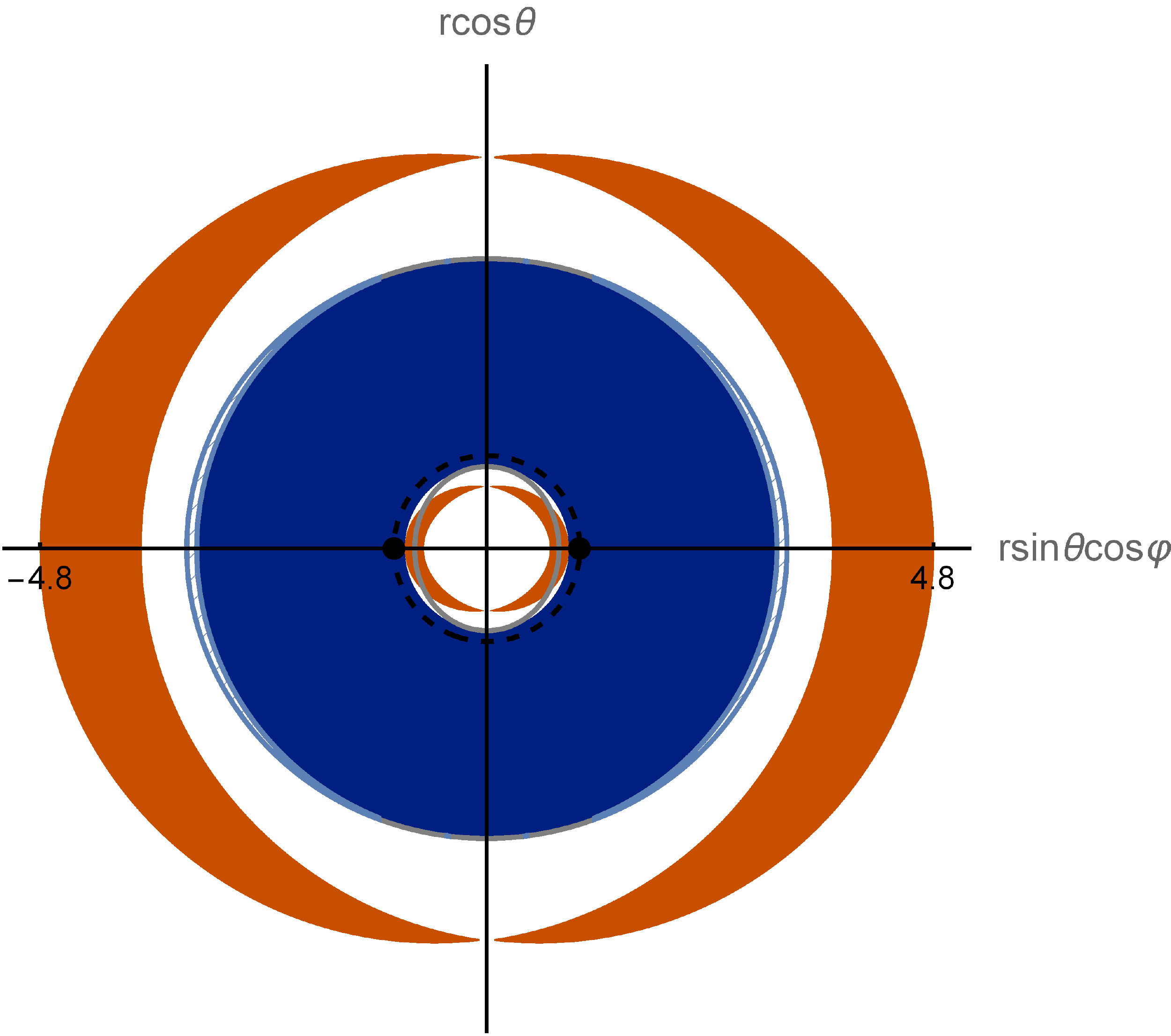}
\includegraphics[width=1\textwidth]{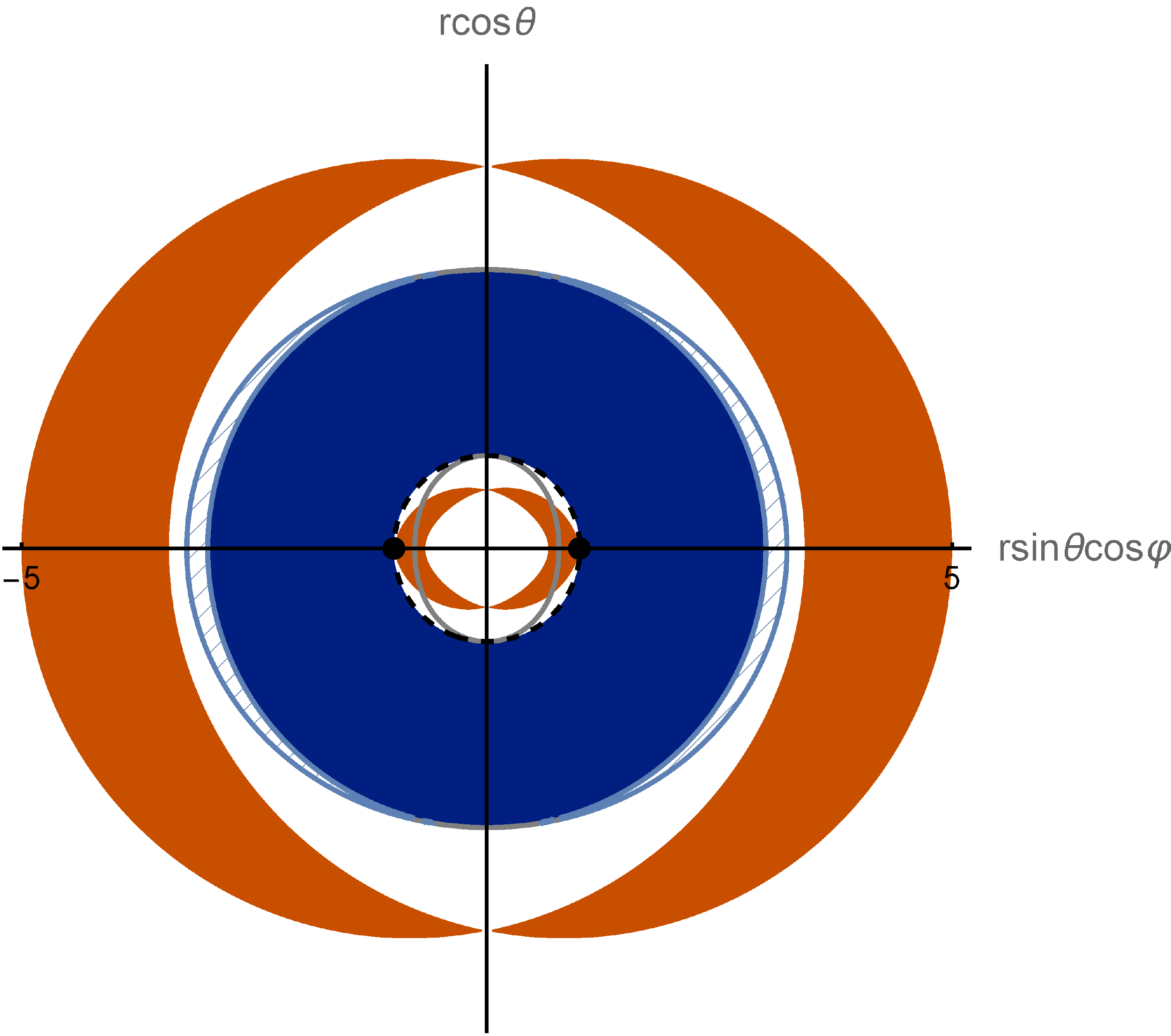}
\centerline{(b)$\beta$=-0.5,q=0}
\end{minipage}
  	\caption{The shapes of the photon regions for different values of $\beta$. The first line, $a=0.5$. The second line, $a=a_{Max}$.}
\label{photon-2}
\end{figure}

\begin{figure}[htbp]
\begin{minipage}[t]{0.45\textwidth}
\includegraphics[width=1\textwidth]{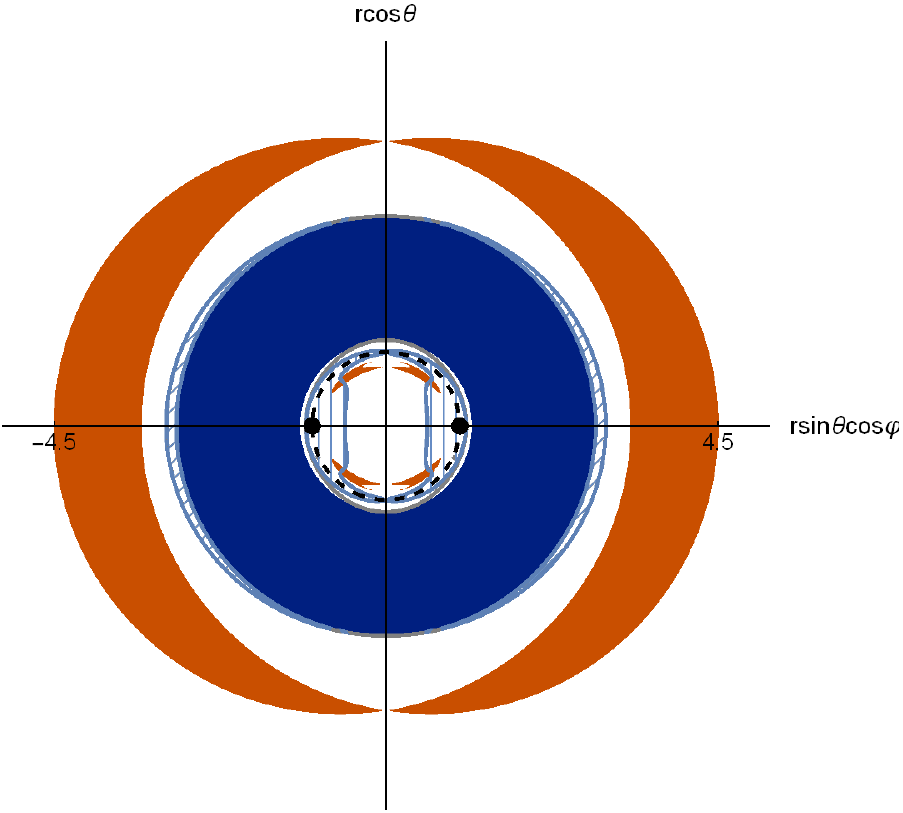}
\includegraphics[width=1\textwidth]{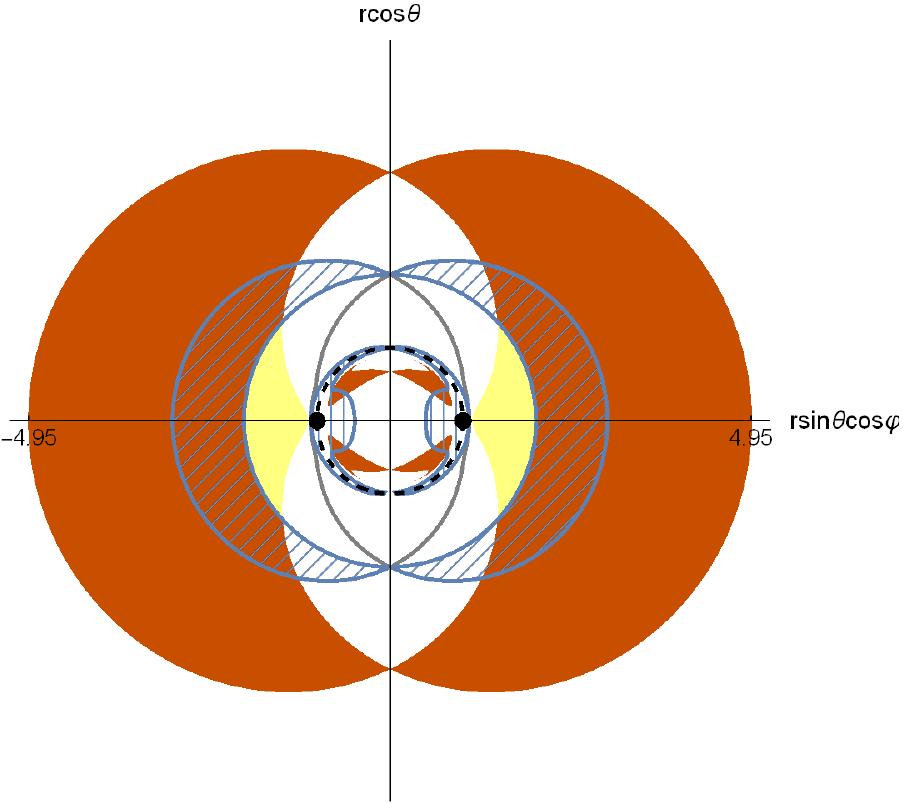}
\centerline{(a)$\beta$=0,q=0.2}
\end{minipage}
\begin{minipage}[t]{0.45\textwidth}
\includegraphics[width=1\textwidth]{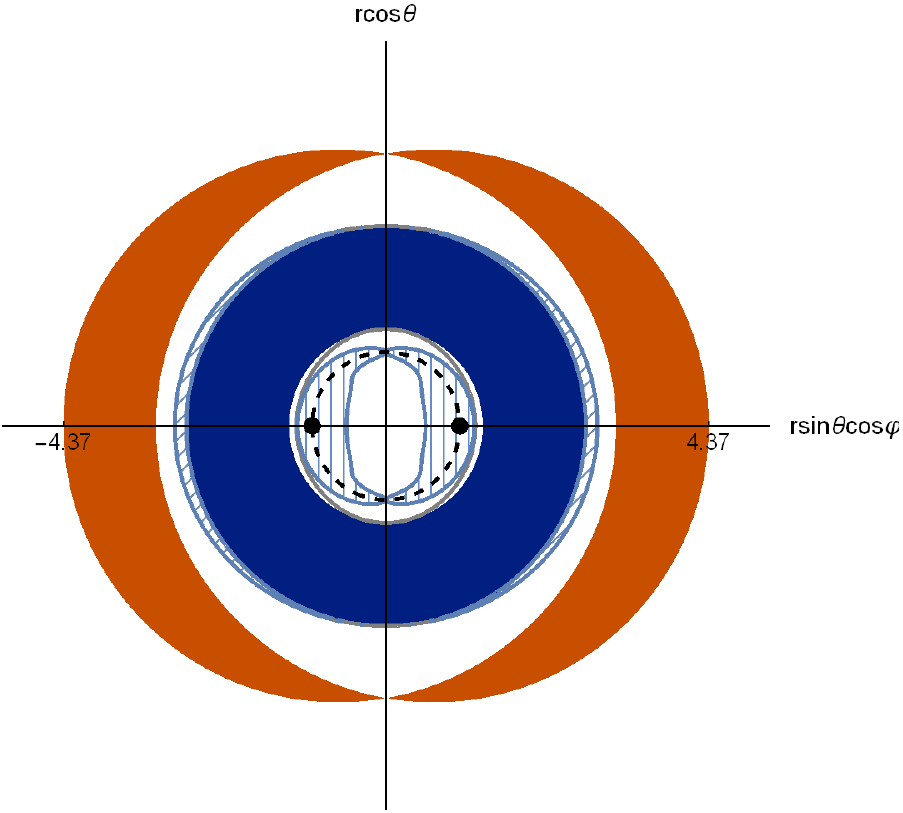}
\includegraphics[width=1\textwidth]{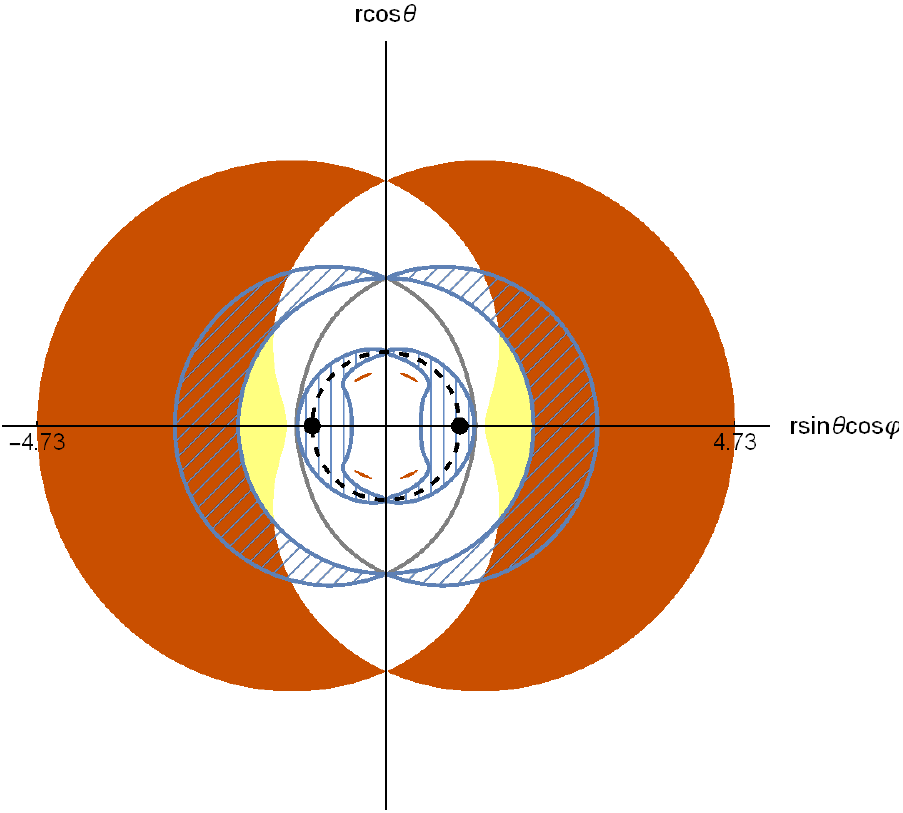}
\centerline{(b)$\beta$=0,q=0.5}
\end{minipage}
  	\caption{The shapes of the photon regions for different values of $q$. The first line, $a=0.5$. The second line, $a=a_{Max}$.}
\label{photon-3}
\end{figure}

When drawing according to the above formula, we used the coordinate transformation in \cite{o1995geometry}, so that the entire space-time can be displayed. And we show the photon regions of the topologically charged rotating black hole in Fig. \ref{photon-1}, Fig. \ref{photon-2} and Fig. \ref{photon-3}. The meaning of the different areas in these images is explained in Fig. \ref{bq}.

In Fig. \ref{photon-1}, we show the effect on the photon regions when the parameter $a$ takes different values. Due to coordinate transformation, the photon regions are divided into two parts, interior photon region ($r<r_{-}$) and exterior photon region ($r>r_{+}$), and they are symmetrical about the coordinate axis. It can be seen that when the value of $a$ is very small, the photon regions degenerate into photon spheres. If the parameters $\beta$ and $q$ are zero at this time, the image describes the Schwarzschild case, and the radius of the photon spheres is $r=3M$, that is, the coordinates marked on the figure are $(\pm4, 0)$. 

\begin{figure}[tbp]%%% Legend
	\centering
	\begin{tabular}{cl}
		\rectc{blau1}  &\large{region with $\Delta \leq 0$}  \\
		\rectc{rot}    &\large{unstable spherical light-rays in $\mathscr{K}$}  \\
		\rectc{orange} &\large{stable spherical light-rays in $\mathscr{K}$}  \\
		\rectl{blau2}  &\large{region with $g_{\phi \phi}<0$ (causality violation)}  \\
		\rectr{gray}   &\large{region with $g_{tt}>0$ (ergosphere)}  \\
		\circpdash{black,dashed}  &\large{throats at $r=0$}  \\
		\textbullet    &\large{ring singularity}
	\end{tabular}
	\caption{Legend for Figs.\ref{photon-1}, \ref{photon-2} and \ref{photon-3} }
	\label{bq}
\end{figure}

When $a\neq0$, the exterior photon region gradually changes from a spherical shape to a crescent shape, and grows with increasing $a$. The interior photon region consists of two connected components that are separated by the ring singularity. Unlike the case where there is only unstable spherical light-rays in exterior photon region, there are two regions of stable and unstable in interior photon region at the same time. With $a$ increased, the distance between the internal and external horizon gradually decreases. When $\beta=q=0$, the shape of the photon region is back to Kerr case. If $\beta$ or $q$ is not zero, the shape of the region with $g_{\phi\phi}<0$ will be significantly deformed. In Figs. \ref{photon-2} and \ref{photon-3}, we show the influence of parameters $\beta$ and $q$ on the shape of various regions, respectively. As shown in Fig. \ref{photon-2}, when $\beta\ \textless0$, the shape of the region with $g_{\phi\phi}<0$ will disappear. And in Fig. \ref{photon-3}, as $q$ increases, the deformation becomes more obvious.

\section{shadow of black hole}\label{5}
\subsection{The shadow shape}
  As mentioned earlier, if a black hole located between a light source and an observer, due to the strong gravity of the black hole, the light path will no longer be a straight line. There will be three situations for this impact, when a photon is emitted from a light source at infinity and bypasses the black hole. First, with a large enough orbital angular momentum, the photon will turn around and reach the observer at infinity. Second, the orbital angular momentum of the photon is too small to resist the gravity and the photon will fall into the black hole, thus the observer cannot see the light source. If there are countless evenly distributed light sources at infinity, because of the second case, the observer will see a dark zone, which is the shadow of the black hole. Third, the orbital angular momentum is in a critical state, which makes the photon always move around the black hole instead of escaping or falling into the black hole. This critical state will help us define the boundary of the shadow. 
\par
For depicting the shadows, we use a stereographic projection from the celestial
sphere onto to a plane with the Cartesian coordinates \cite{wang2019shadows}:
  \begin{equation}\label{eq:xy}
  \begin{split}
  x&=-\xi\csc{\theta},\\
  y&=\pm\sqrt{\eta+a^2\cos^2{\theta}-\xi^2\cot^2{\theta}},
  \end{split}
  \end{equation}
  the angle between the line of sight of observer and the black hole's rotation axis is $\theta$.
\par
\begin{figure}[htbp]
  	\centering
  	\subfigure[~]{
  		\includegraphics[width=.47\textwidth]{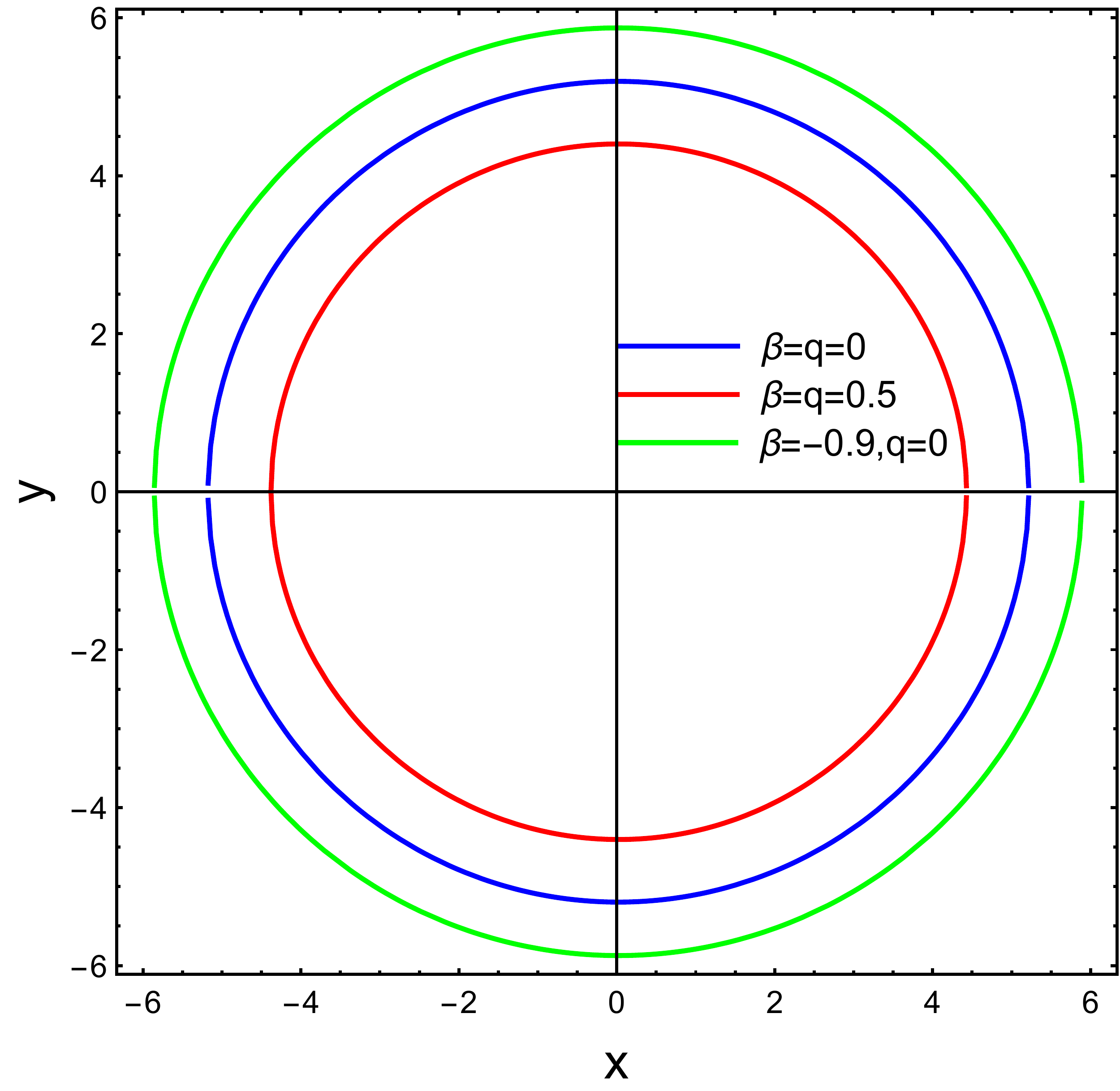}
       \label{fig:shadow-11}}
   \subfigure[~]{
  		\includegraphics[width=.47\textwidth]{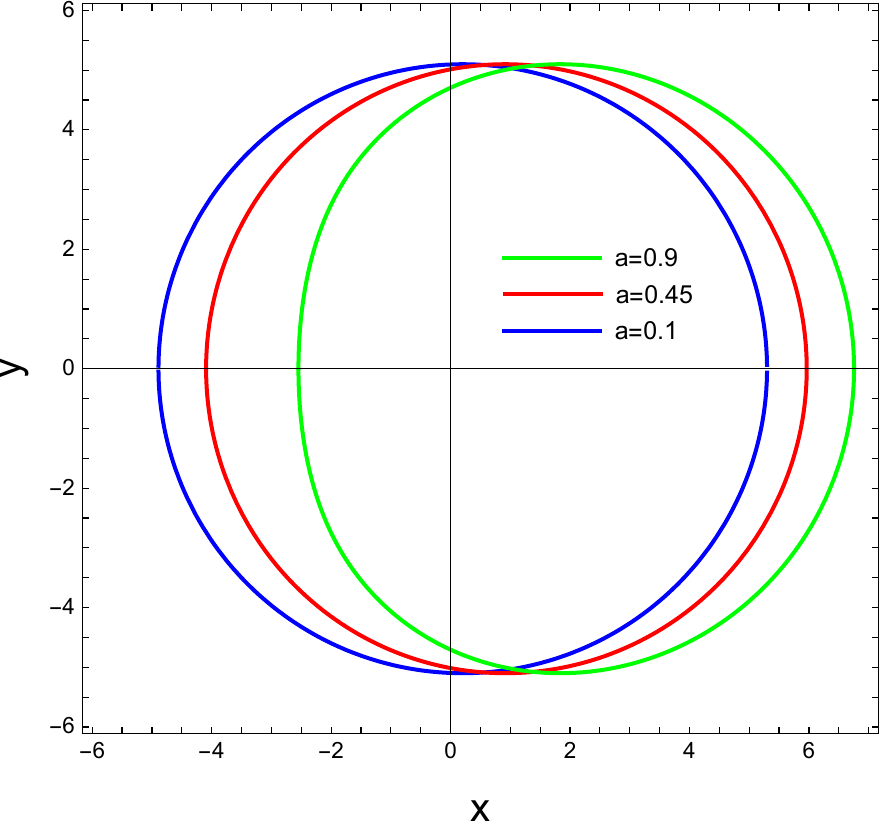}
  	\label{fig:shadow-12}}
 \caption{Shadows cast by black holes. (a) The non-rotating case ($a=0$), $\theta=\pi/2$. (b)$a\neq0$, $\beta=q=0$, $\theta=\pi/2$.}
\label{fig:shadow-1}
  \end{figure}

Now, we show the shadow of a topologically charged rotating black hole using Eq.~(\ref{eq:xy}). In Fig. \ref{fig:shadow-1}, we show the different cases of the shadow of the black hole with and without rotation parameter $a$. And in Fig. \ref{fig:shadow-11}, it is easily to see that the shape of non-rotating black hole is a standard circle, while the size of the radius is related to the parameters $\beta$ and $q$, and with increasing the value of the parameters $\beta$ and $q$ the size of shadow is decreasing. If $\beta$ takes a negative value, the size of the shadow will become larger than that of Schwarzschild case, which can be clearly seen from the figure. When spin parameter $a\neq0$, the rotating black hole will have a dragging effect on the photon, which will cause the shape of shadow to deform in the direction perpendicular to the axis of rotation. As shown in Fig. \ref{fig:shadow-12}, fixed the parameters $\beta$ and $q$, while changing the spin parameter $a$ from 0.1 to 0.9. As the value of $a$ increases, the deformation of the shape of shadow is gradually obvious.
\par
\begin{figure}[htbp]
  	\subfigure[~]{
  		\includegraphics[width=.47\textwidth]{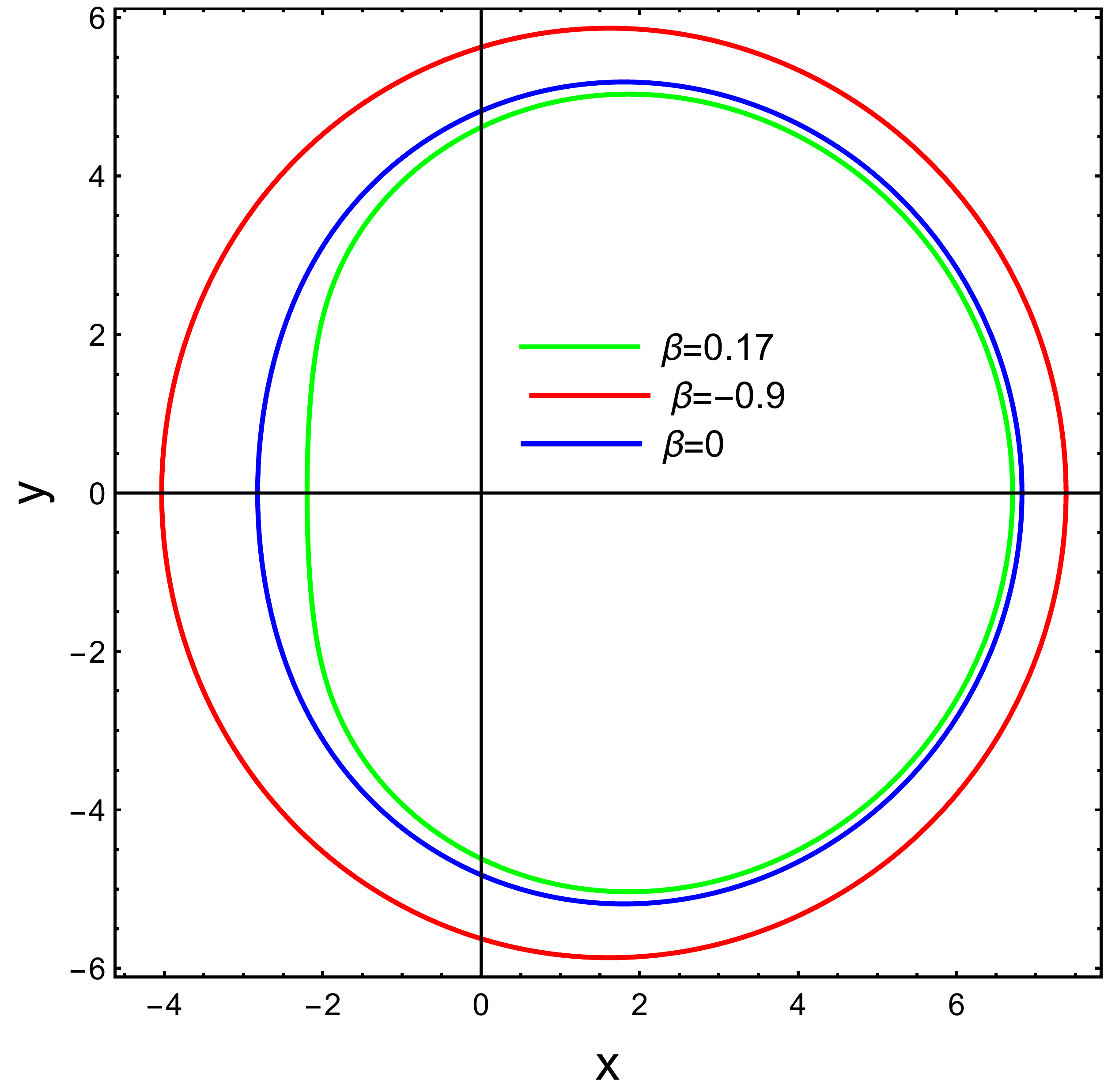}}
  	\subfigure[~]{
  		\includegraphics[width=.47\textwidth]{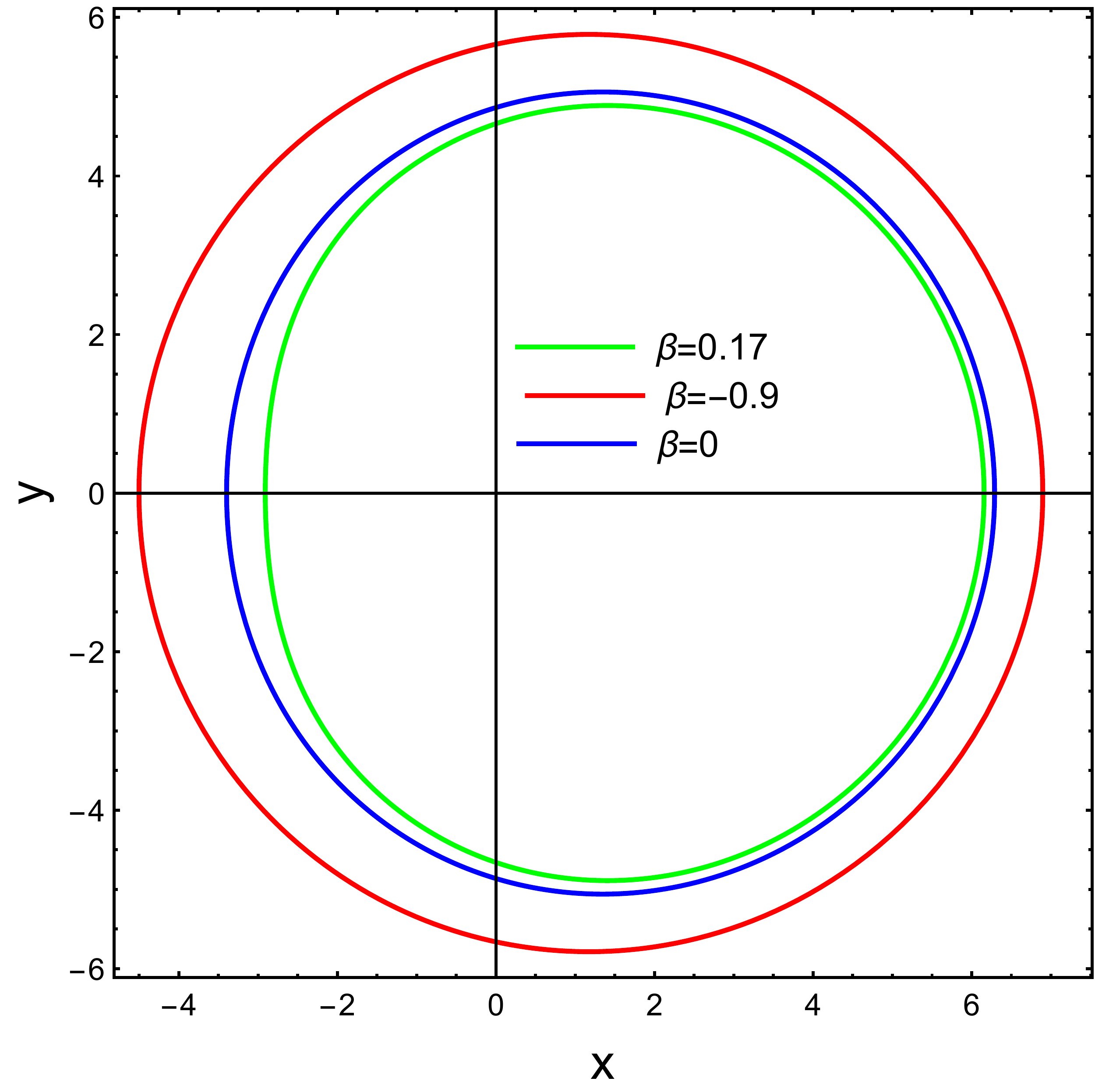}}
  	\subfigure[~]{
  		\includegraphics[width=.47\textwidth]{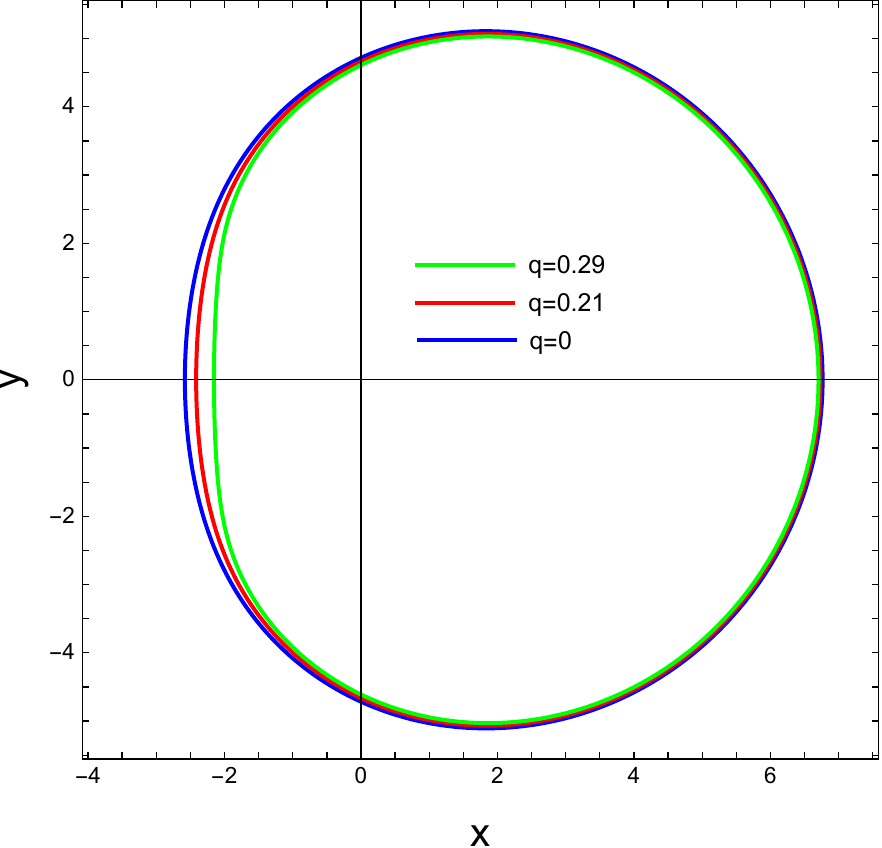}}
  	\subfigure[~]{
  	    \includegraphics[width=.47\textwidth]{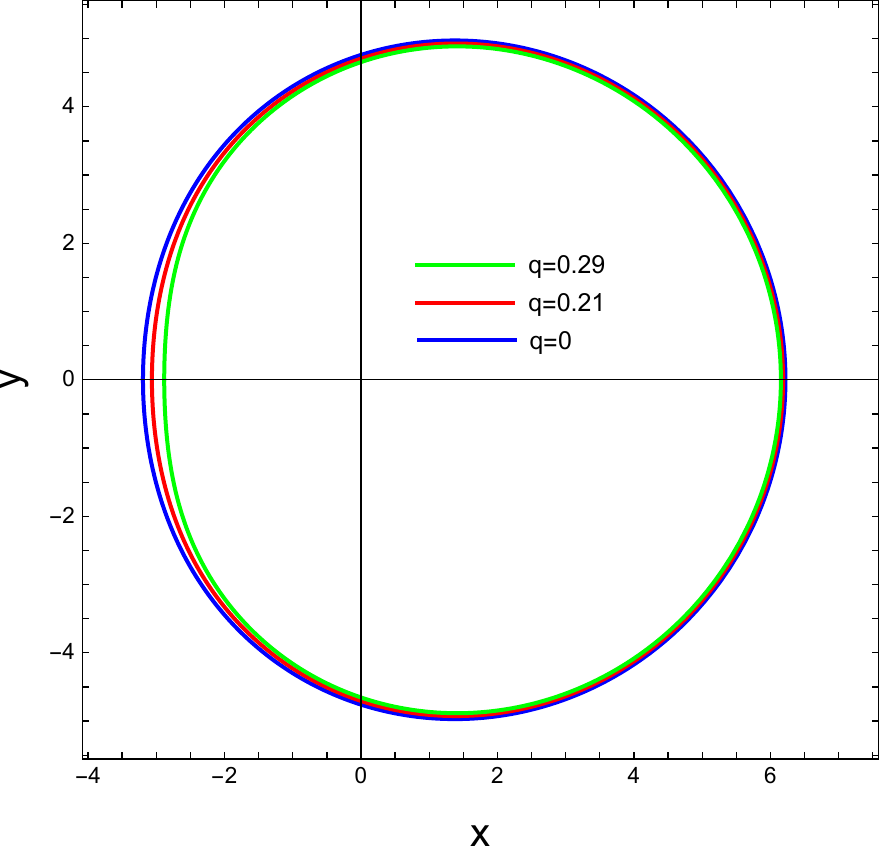}}
  	\caption{The shapes of the black hole shadow for different values of the parameters. (a) $a=0.9,q=0.1,\theta=\pi/2$. (b) $a=0.9,q=0.1,\theta=\pi/4$. (c) $a=0.9,\beta=0.1,\theta=\pi/2$. (d) $a=0.9,\beta=0.1,\theta=\pi/4$.}
  	\label{fig:shadow-2}
  \end{figure}

At the same time, we show the shapes of the shadow  for different values of $\beta$, $q$ and the inclination angle $\theta$ while fixing $a$ in Fig. \ref{fig:shadow-2}. Similar to changing $a$, the increase of $\beta$ and $q$ will increase the deformation. However, a parameter $\beta$ less than zero will not only enlarge the shadow, but also inhibit the deformation of the shadow, this phenomenon can be seen more clearly in the next section.

  \begin{figure}[htbp]
  	\centering
  	\subfigure{
  		\includegraphics[width=.60\textwidth]{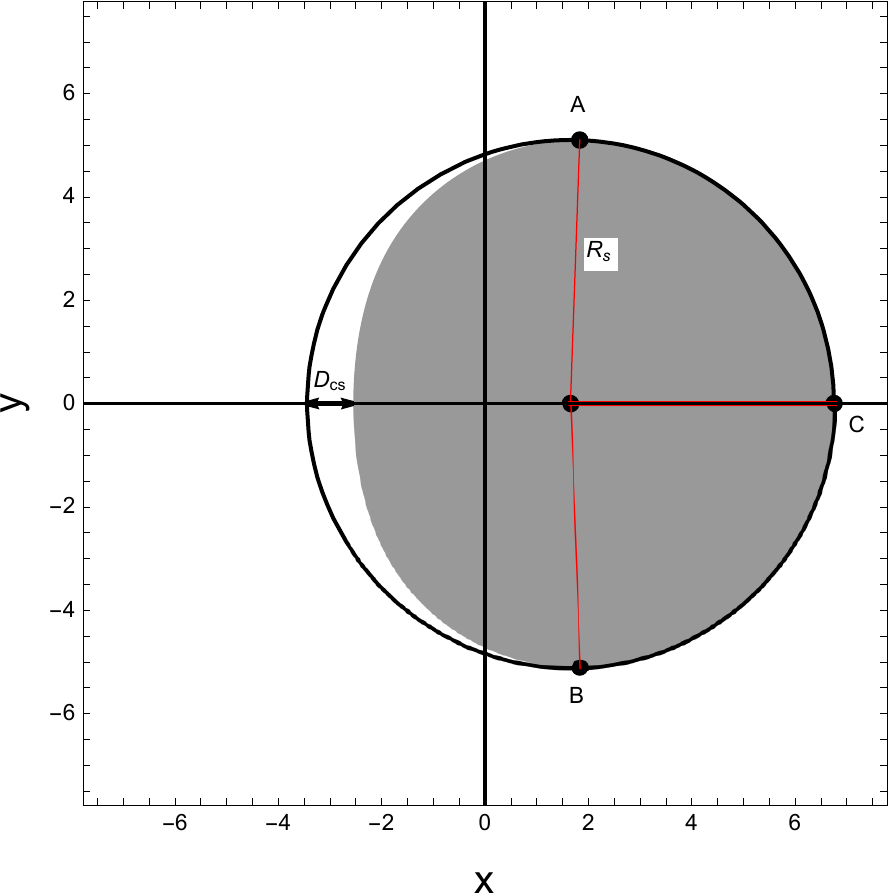}
}
  	\caption{The observable parameter, the radius $R_{s}$(red line), and the distortion parameter $\delta_{s}=D_{cs}/R_{s}$ are described as the apparent shape of the black hole.}
  	\label{fig:shadow-define}
  \end{figure}

  	  \begin{figure}[htbp]
      \subfigure[~]{
  		\includegraphics[width=.47\textwidth]{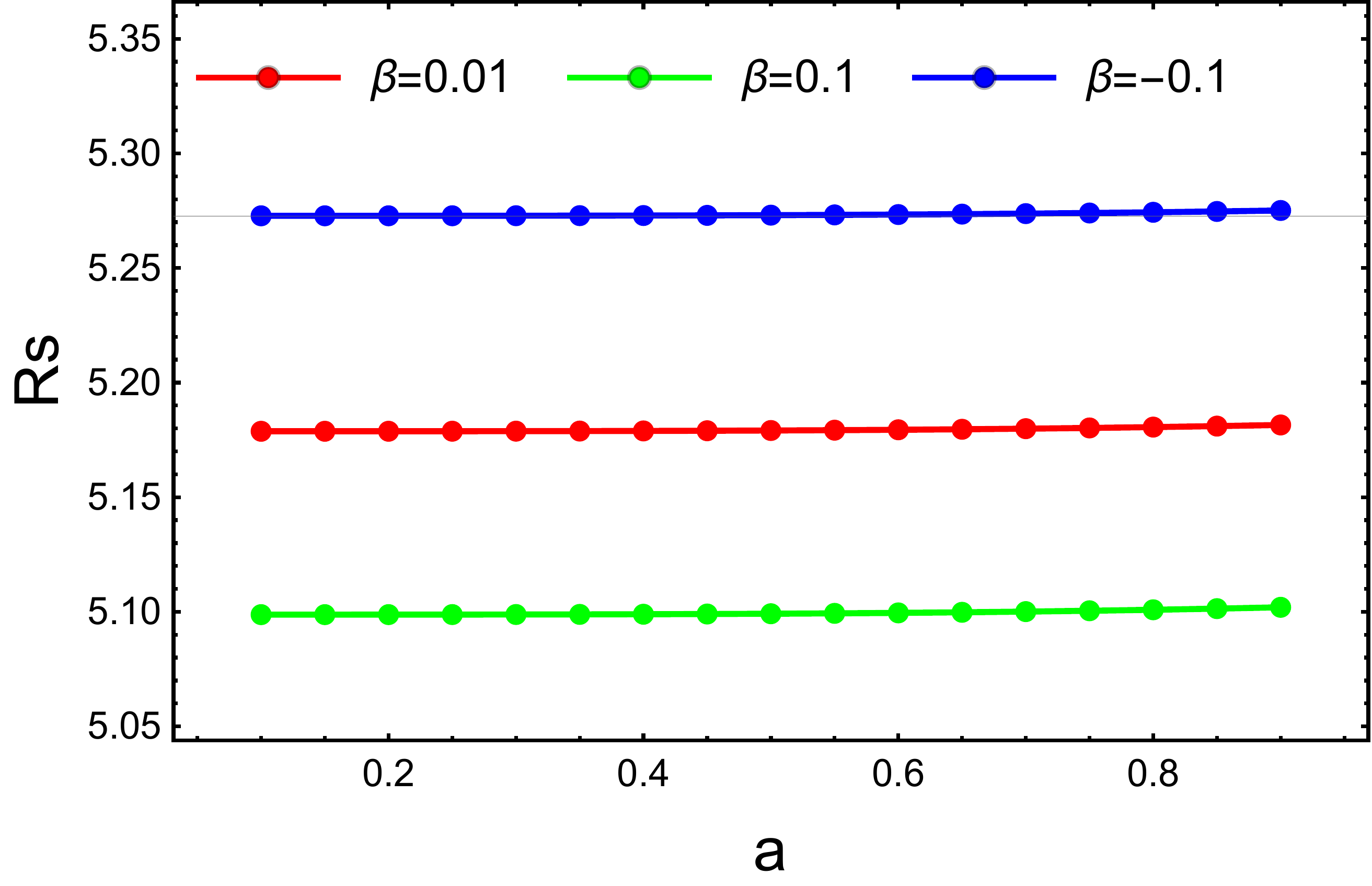}
  		\label{fig:Rs-a}
  	}
  	\subfigure[~]{
  		\includegraphics[width=.47\textwidth]{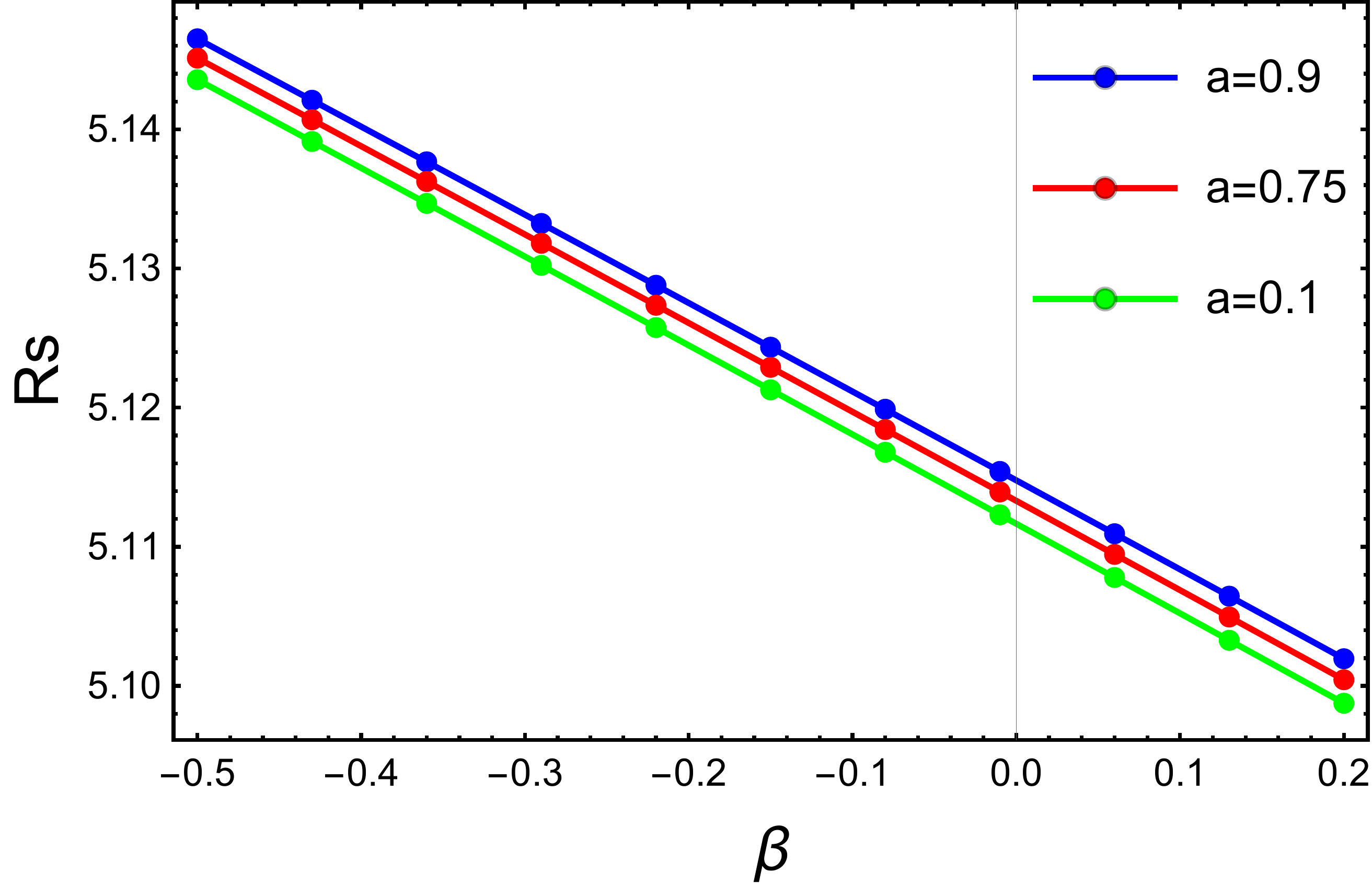}
  		\label{fig:Rs-b}
  	}
   \subfigure[~]{
  		\includegraphics[width=.47\textwidth]{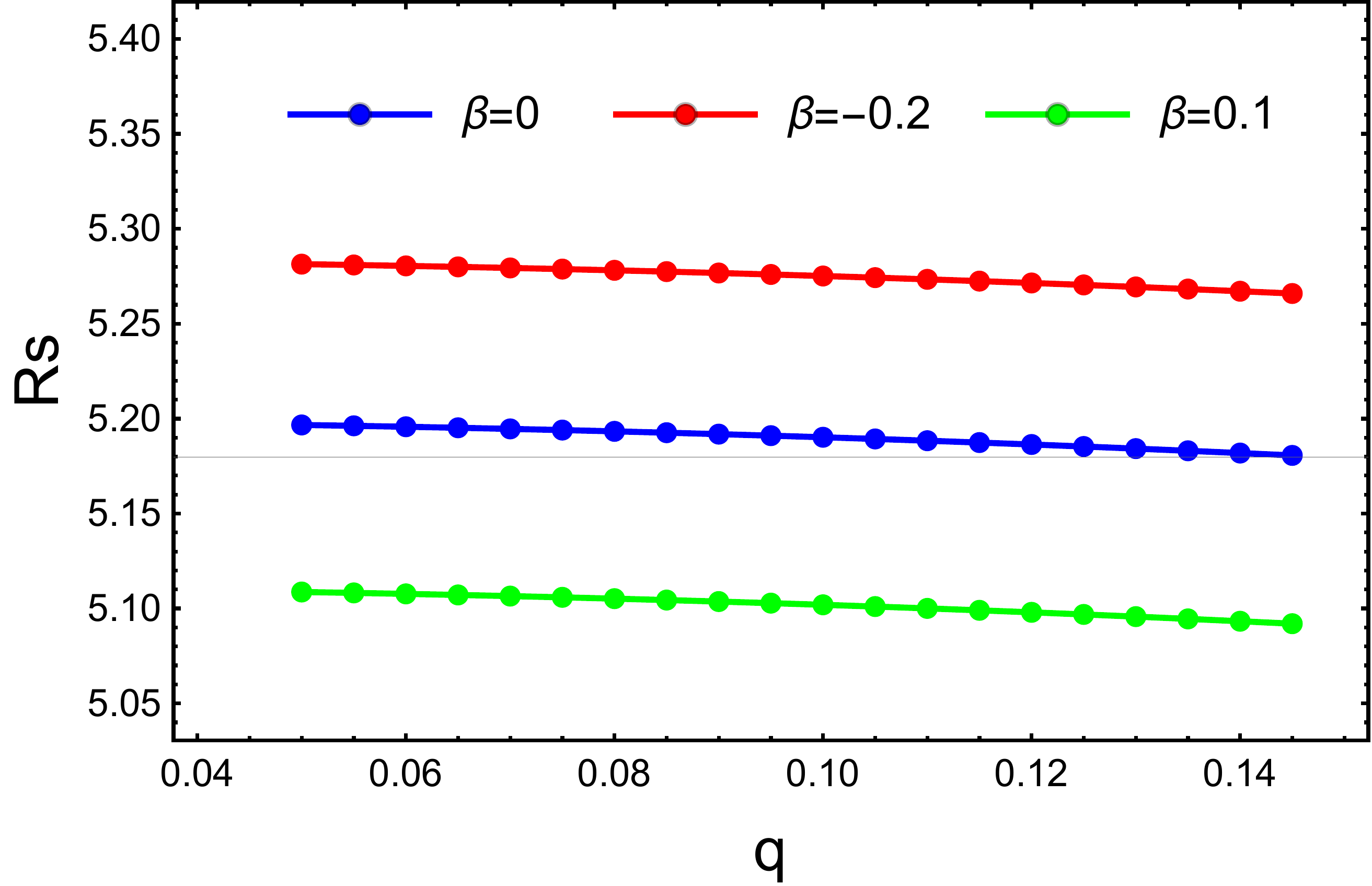}
  		\label{fig:Rs-q}
  	}
  	\caption{(a) The radius $R_{s}$ of the black hole shadow against $a$ for different values of parameter $\beta$, $q=0.1$. (b) $R_{s}$ against $\beta$ for different values of parameter $a$, $q=0.1$. (c) $R_{s}$ against $q$ for different values of parameter $\beta$, $a=0.9$.}
  	\label{Rs-a-b-q}
  \end{figure}

  \begin{figure}[htbp]
  	\subfigure[~]{
  		\includegraphics[width=.47\textwidth]{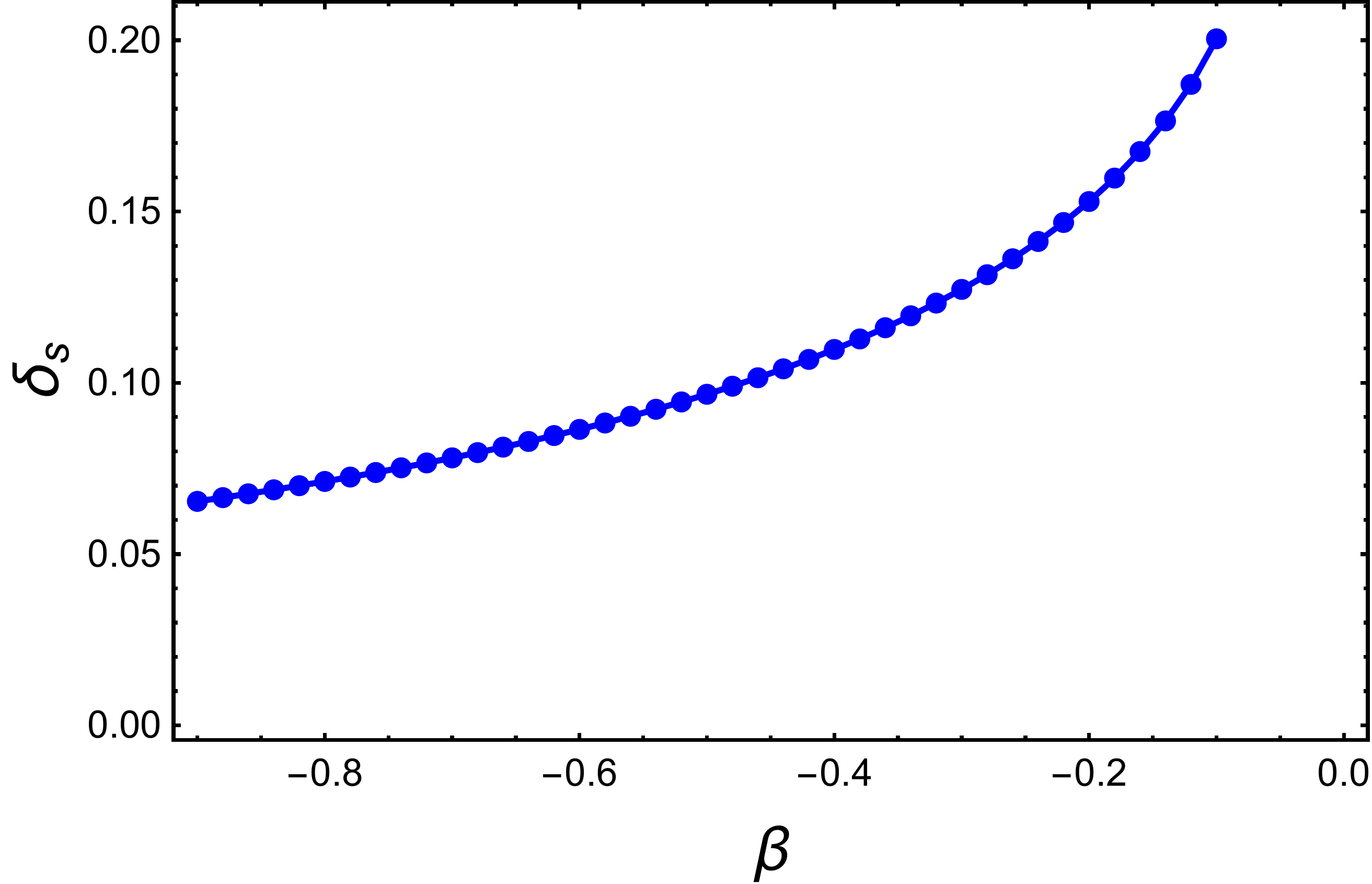}
  		\label{fig:delta-b}
  	}
  	\subfigure[~]{
  		\includegraphics[width=.47\textwidth]{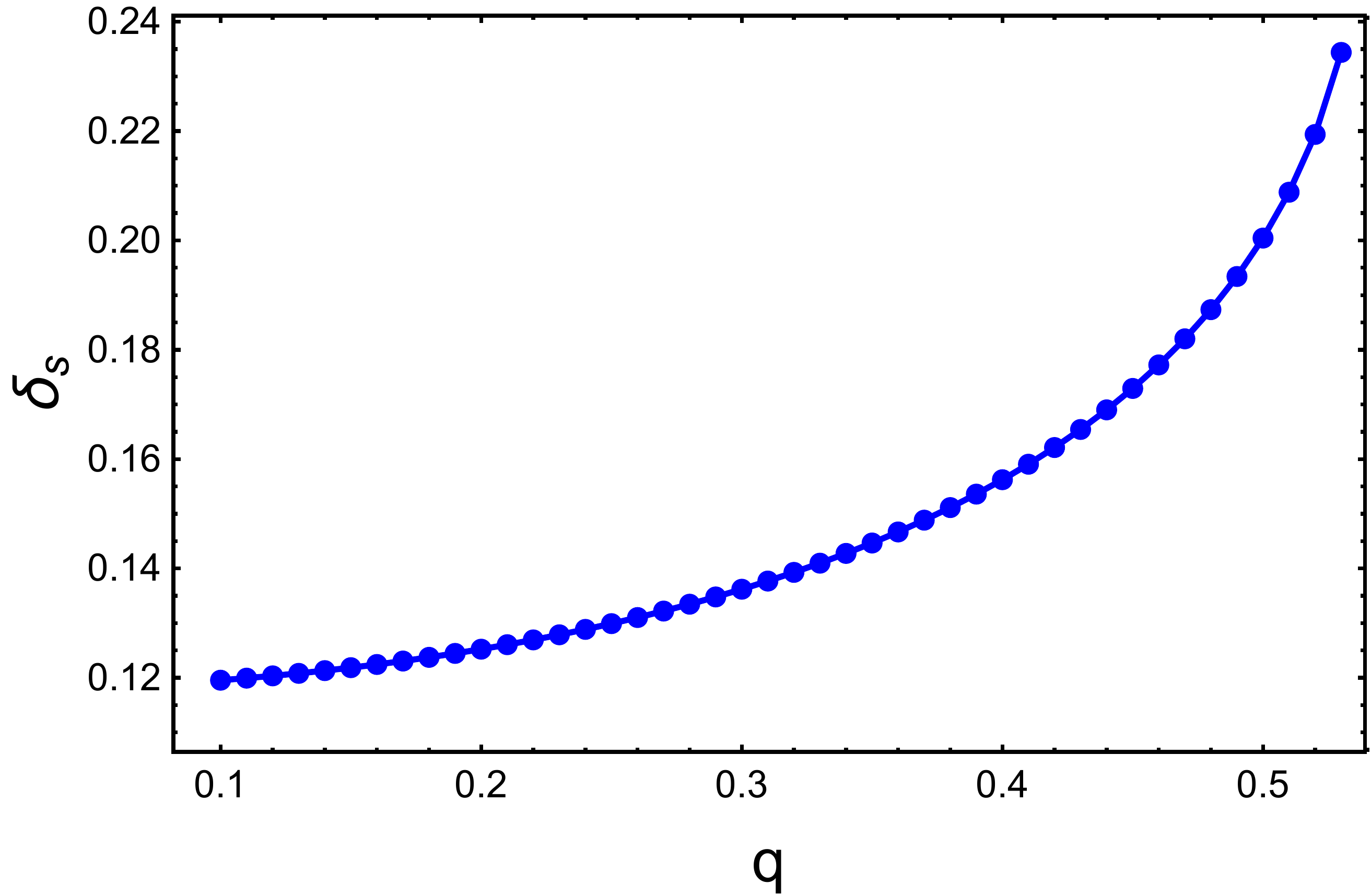}
  		\label{fig:delta-q}
  	}
   \subfigure[~]{
  		\includegraphics[width=.47\textwidth]{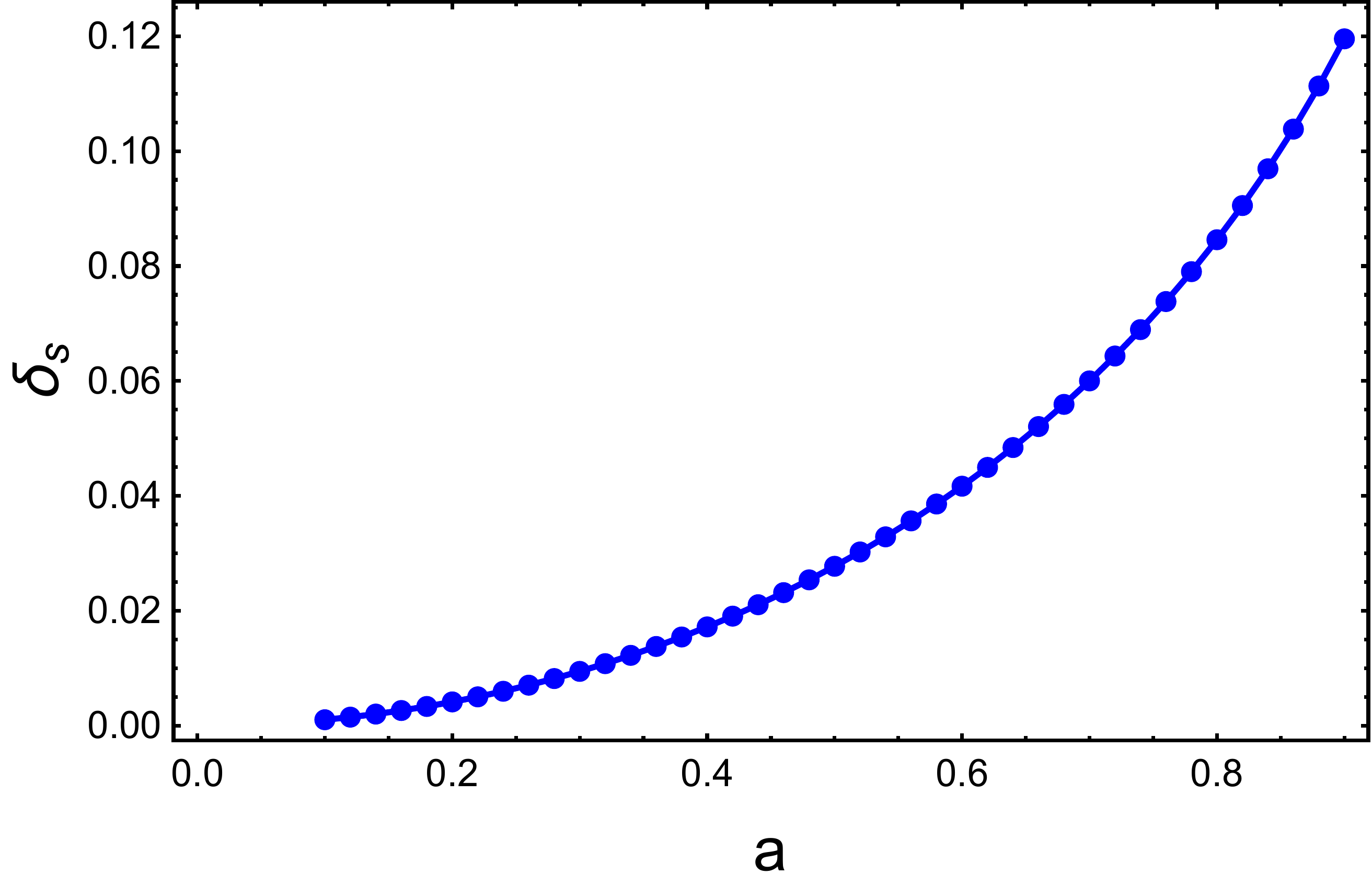}
  		\label{fig:delta-a}
  	}
  	\caption{The distortion $\delta_{s}$ of the black hole shadow for different values of the parameters. (a)$a=0.9$, $q=0.1$. (b)$a=0.9$, $\beta=-0.1$. (c)$\beta=q=0.1$.}
  	\label{delta-a-b-q}
  \end{figure}

  \begin{figure}[htbp]
  	\subfigure[~]{
  		\includegraphics[width=.47\textwidth]{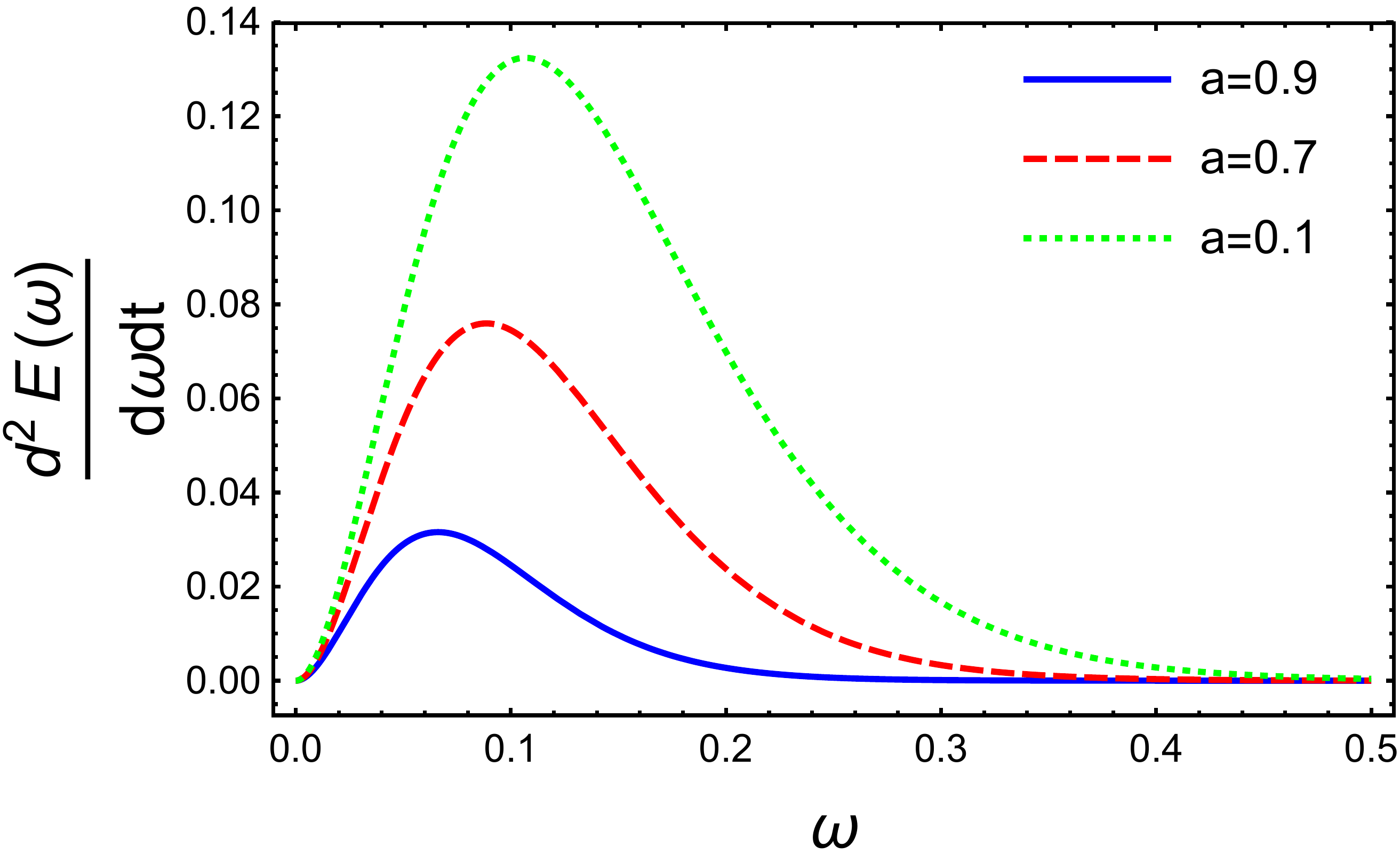}
  		\label{fig:E-a}
  	}
  	\subfigure[~]{
  		\includegraphics[width=.47\textwidth]{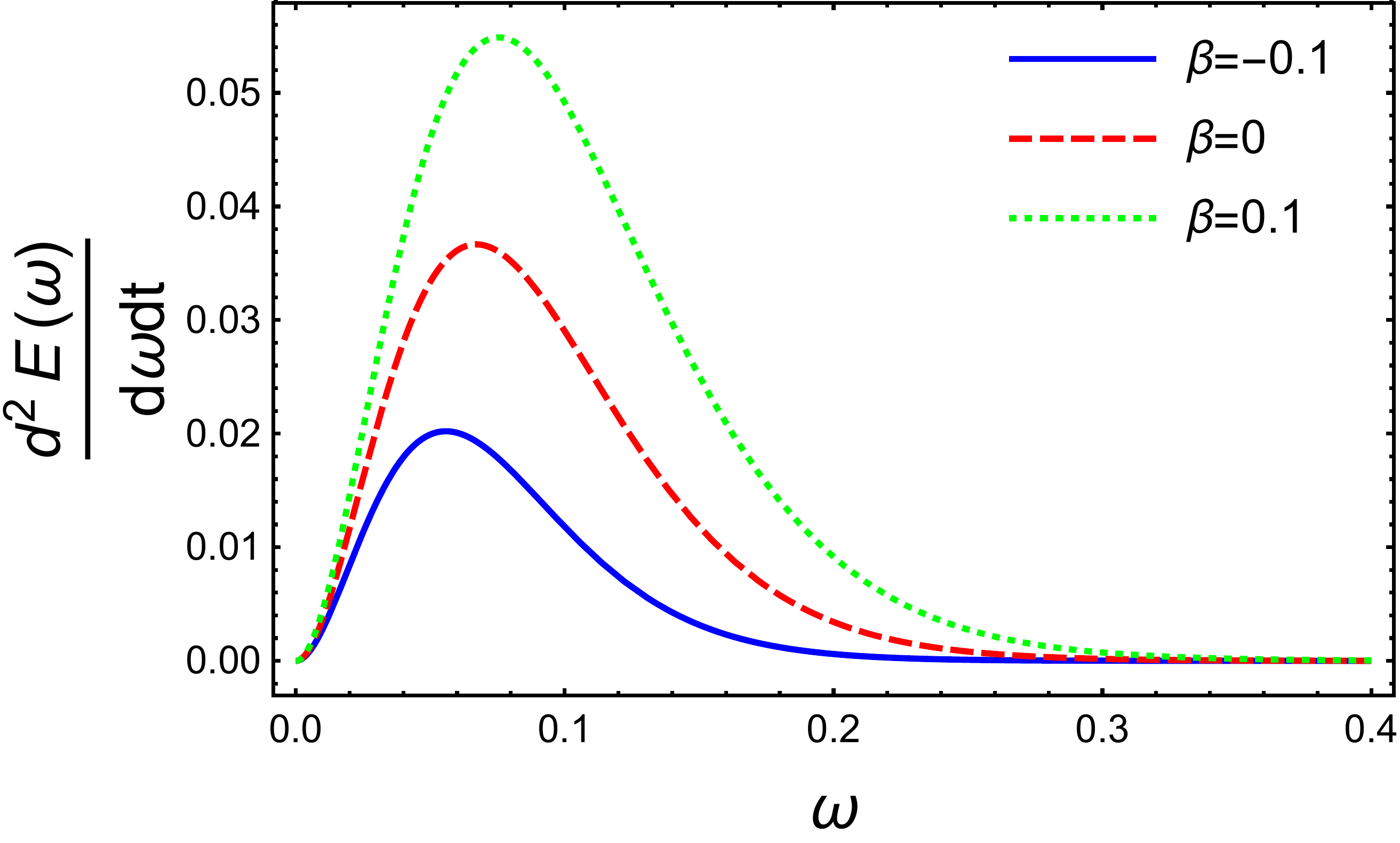}
  		\label{fig:E-b}
  	}
   \subfigure[~]{
  		\includegraphics[width=.47\textwidth]{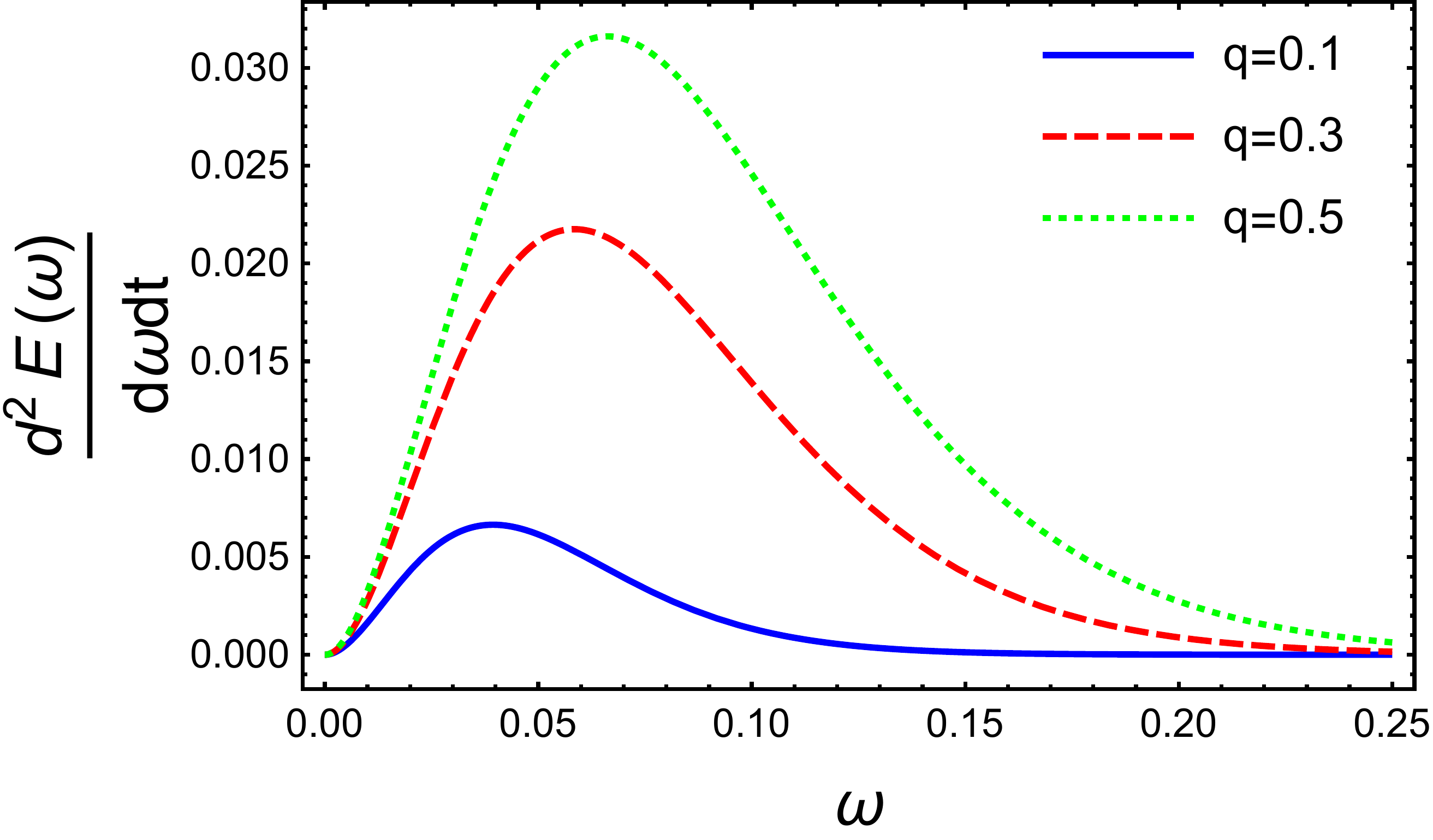}
  		\label{fig:E-q}
  	}
  	\caption{Behaviors of the energy emission rate for different values of parameters. (a)$\beta=-0.1,q=0.1$. (b)$a=0.9$, $q=0.1$. (c)$a=0.9$, $\beta=-0.1$.}
  	\label{E-a-b-q}
  \end{figure}

\subsection{Observables}
  In order to characterize the shadow of a black hole, it is useful to introduce two observables the radius $R_{s}$ and the distortion parameter $\delta_{s}$ that approximately characterize its shape. As shown in Fig. \ref{fig:shadow-define}, the shape of the shadow is a deformed circle, so we select three specific points on the image: the top point $A(x_{t},y_{t})$, the bottom point $B(x_{b},y_{b})$, the right point $C(x_{r},0)$, and draw a standard circle covering the shadow. The radius $R_{s}$ of the shadow is hereby defined by the radius of the circle. The distortion parameter $\delta_{s}$ is defined as $\delta_{s}=D_{cs}/R_{s}$, where $D_{cs}$ is the difference between the right end points of the circle and of the shadow. The observable $R_{s}$ is defined as
  \begin{equation}\label{eq:RS}
  R_{s}=\frac{\left(x_{t}-x_{r}\right)^2+y^2_{t}}{2\left(x_{r}-x_{t}\right)}.
  \end{equation}
\par
We numerically calculate these observables, and the radius $R_{s}$ is clearly shown in Fig. \ref{Rs-a-b-q}. In Fig. \ref{fig:Rs-a}, we can find that for fixed $q$, $R_{s}$ increases with the spin parameter $a$ increases and decreases as the tidal charge $\beta$ increase. Fig. \ref{fig:Rs-b} further shows this trend. This is because positive parameter $\beta$ makes the shadow shape smaller, and parameter $a$ increases the deformation of the shadow shape as shown above. At the same time, we show that for fixed a, $R_{s}$ will also decreases as the electric charge parameter $q$ increases, just like $\beta$. Interestingly, the result shows that the shadow of a topologically charged rotating black hole is smaller.
\par
The distortion parameter $\delta_{s}$ are clearly shown in Fig. \ref{delta-a-b-q}. As shown in Fig. \ref{fig:delta-a}, when the parameters $\beta$ and $q$ are fixed, the shape of the shadow is obviously deformed with the increase of $a$. In Fig. \ref{fig:delta-b} and Fig. \ref{fig:delta-q}, after parameter $a$ is fixed, the increase of $\beta$ and $q$ will exacerbate the deformation of the shadow.

\par
 From \cite{wei2013observing}, the authors proposed that the area of the black hole shadow is approximately equal to the high-energy absorption cross-section for an observer at the equatorial plane at infinity. According to this assumption, the energy emission rate of the black hole in high energy case is
  \begin{equation}\label{emission}
 \frac{\mathrm{d}^2 E\left(\omega\right)}{\mathrm{d}\omega\mathrm{d}t}=\frac{2\pi^3R_{s}^2}{e^{\omega/T_{H}}-1}\omega^3, 
 \end{equation}
where $R_{s}$ is given in Eq.~(\ref{eq:RS}), and $T_{H}$ is the Hawking temperature, which in \cite{larranaga2013topologically} given by
\begin{equation}\label{eq:T}
T=\frac{r_{+}}{4\pi\left(r_{+}^2+a^2\right)}\left(1-\frac{a^2}{r_{+}^2}-\frac{\beta+q^2}{r_{+}^2}-\frac{1}{48} \frac{\kappa_{5}^4 q^4}{r_{+}^6}\right),
\end{equation}
then, we describe the energy emission rate against the frequency $\omega$ for different parameters respectively, a, $\beta$ and q,  in Fig. \ref{E-a-b-q}.  For each parameter, the curve has a peak. And the peak will decrease as the corresponding parameter increases, while the other two parameters keep fixed.

\section{Conclusions and discussions}\label{6}
In this paper, we have discussed the shadow of a topologically charged rotating black hole. Through this metrics, we have discussed horizon of the black hole. It can be seen that as the parameter increases, the distance between two horizons will gradually decrease. 

Then we studied the photon region, which is a standard circle in the Schwarzschild case, and splits in the rotating case. This is also the reason for the deformation of the shadow of the rotating black hole. Interestingly, if the charge parameter $q$ is not zero, the causality violation region will be deformed, at the same time if $\beta$ is negative, the region will disappear. Further, we draw the shadow of the black hole in the two cases of non-rotation and rotation. Similar to the conclusion that the black hole horizon changes with parameters, the increase of the parameters $\beta$ and $q$ will make the shape of the black hole shadow smaller, but when $\beta$ takes a negative value, the shadow will increase. 

At last, we studied two observables, the radius $R_{s}$, distortions $\delta_{s}$. As shown in the figure, $R_{s}$ increases with the increase of three parameters $a$, $\beta$ and $q$. In addition, suppose the area of the black hole shadow is equal to the high-energy absorption cross section, based on this assumption, we studied the energy emission rate. From the figures, one can find that with the increase of parameters $a$, $\beta$ and $q$, the peak decreases and moves to the low frequency.

It is currently believed that there are supermassive black holes in the centers of many galaxies. Therefore, the investigation of the shadow may be a very useful tool to study the nature of the black hole. 

\section*{Conflicts of Interest}
  The authors declare that there are no conflicts of interest regarding the publication of this paper.

\section*{Acknowledgments}
  We would like to thank the National Natural Science Foundation of China (Grant No.11571342) for supporting us on this work.

\section*{References}

 \bibliographystyle{unsrt}
 \bibliography{ref}
\end{document}